\documentclass[12pt]{article}

\begin{document}
\begin{center}
{\bf EFFECTIVE ACTIONS, RADII AND ELECTROMAGNETIC POLARIZABILITIES
OF HADRONS IN QCD STRING THEORY}\\
\vspace{5mm}
 S.I. Kruglov \footnote{E-mail:
skruglov23@hotmail.com}\\
\vspace{5mm}
\textit{International Educational Centre, 2727 Steeles Ave. W, \# 202, \\
Toronto, Ontario, Canada M3J 3G9}
\end{center}

\begin{abstract}
A nonperturbative approach to QCD describing confinement and
chiral symmetry breaking is discussed. It is based on the path
integral representation of Green's function of quarks and leads to
the QCD string theory. The effective actions for mesons and
baryons in the external uniform static electromagnetic fields are
obtained. The area law of the Wilson loop integral, the
approximation of the Nambu-Goto straight-line string, and the
asymmetric quark-diquark structure of nucleons are used to
simplify the problem. The spin-orbit and spin-spin interactions of
quarks are treated as a perturbation. Using the virial theorem we
estimate the mean radii of hadrons in terms of the string tension
and the Airy function zeros. On the basis of the perturbation
theory in small external electromagnetic fields we derive the
electromagnetic polarizabilities of nucleons. The electric and
diamagnetic polarizabilities of a proton are $\bar{\alpha}_p=
10\times 10^{-4}~ fm^3$, $\beta _p^{dia}=-8\times 10^{-4}~ fm^3$
and for a neutron we find $\bar{\alpha }_n=4.2\times 10^{-4}~ fm^3
$, $\beta _n^{dia}= -5.4\times 10^{-4}~ fm^3$. Using the $\Delta$
contribution to the paramagnetic polarizability of the nucleons,
reasonable values of the magnetic polarizabilities
$\bar{\beta}_p=(5\pm 3)\times 10^{-4}~ fm^3$,
$\bar{\beta}_n=(7.6\pm 3)\times 10^{-4}~ fm^3$ are estimated.
\end{abstract}

\section{Introduction}

It is now understood that Quantum Chromodynamics (QCD) (see [1]) really describes
the strong interactions of hadrons. However, QCD deals with nonobserved objects:
quarks and gluons. The standard perturbation theory in the small QCD coupling
constant ($\alpha_s$) uses Feynmann diagrams and is applicable in the ultraviolet region.
In the infrared region, which is important for hadron physics, the coupling constant
is large, and it is impossible to apply perturbation technique. In this case
the nonperturbative methods should be used. One of the powerful methods is lattice
simulations of QCD giving data about hadron spectrum, matrix elements, and
scattering. The second way is to develop analytical nonperturbative methods
which allow us to understand deeper nonperturbative phenomena. There are some
phenomenological approaches for describing hadron physics such as potential models [2],
method of the effective Lagrangians [3], the QCD sum rule approach [4] and others.
However, these schemes do not deal with the fundamental QCD Lagrangian.

One of the important problems of particle physics in the infrared region is the
confinement of quarks. According to this phenomenon only colourless objects
(hadrons) are observable. This means that quarks and gluons which are present in
the QCD Lagrangian are absent in the physical spectrum. Forces between quarks and
gluons combine them into hadronic colourless states. The appearance of such
forces with a linear potential is explained by the string configurations of gluonic
fields named the QCD string. Mathematically this phenomenon is expressed by the
Wilson area law. Confinement of quarks and gluons can be explained by the dual
Meissner effect [5] in the framework of the Abelian-projected method [6]. In
accordance  with this scenario Abelian magnetic monopoles appear after Abelian
projecting Yang-Mills gluonic fields. These magnetic monopoles condense to form
vacuum with flux tubes (strings) between quarks. The same picture occurs in the
Ginzburg-Landau theory of superconductivity and its relativistic version -
Abelian Higgs Model. In these theories the magnetic Abrikosov-Nilsen-Olesen strings
[7] appear due to the condensation of Cooper pairs (electrons). Thus, chromoelectric
strings which ensure confinement in the framework of QCD are dual to the
Abrikosov-Nilsen-Olesen strings.

It should be noted that QCD strings are treated as effective and are real gluonic
flux tubes. This is a distinction of QCD strings in $4D$ which can be described by the
Nambu-Goto action [8] from the fundamental strings [9]. Therefore there is no
 problem to quantize QCD strings and we avoid the problem of conformal anomaly
in $D\neq26$. Superstring theories were introduced to unify strong, electroweak,
and gravitational interactions.

The goal of this review is to outline the effective string theory which appears from
QCD in the infrared region, and allow us to make some estimations of electromagnetic
characteristics of hadrons.

There is another nonperturbative phenomenon of QCD - chiral symmetry breaking (CSB) the
QCD Lagrangian with massless quarks. For massless quarks, the QCD Lagrangian possesses
the chiral $U(N_f)\times U(N_f)$ symmetry which is spontaneously broken. As a result
there is no parity degeneracy of hadrons. The real difference of hadron masses
with opposite parity is few hundred $MeV$. The consequence of chiral symmetry breaking
QCD is the presence of the Goldstone bosons in the physical spectrum, which are associated
with $\pi$-mesons.

CSB leads to a nonvanishing quark condensate ($\left\langle \bar
{q}q\right\rangle $) [10]. As a result the light quarks ($u$, $d$ quarks) with
current masses $m_u\simeq m_d\simeq 7~MeV$ acquire the dynamical masses
$\mu _u\simeq \mu _d\simeq 320~MeV$. This phenomenon is important for light
pseudoscalar mesons because they possess the Nambu-Goldstone nature. Nonzero gluon
condensate is responsible for confinement [4].
To obtain low masses of pseudoscalar mesons one needs to take into account the spin
interactions of quarks.

Some models were proposed for the explanation of CSB. In instanton vacuum theory (IVT) [11],
CSB was explained by the nonvanishing density of quark zero modes. This approach
is based on the representation of vacuum as an ensemble of classical instanton-
antiinstanton configurations. To have zero topological charge of the QCD vacuum
the number of instantons should be equal to the number of antiinstantons. However,
a superposition of fields with opposite topological charges is instable as instantons
and antiinstantons can annihilate, and in addition such a superposition of fields
is not a solution of classical equations of motion. Classical configurations
should have high density to reproduce the phenomenological gluon condensate [4],
and can distort the original solutions. Besides, IVT can not explain confinement of
quarks.

In [12] CSB was confirmed on the basis of gluon propagator which is the solution
of Dyson-Schwinger equations.

CSB was also studied in detail in the framework of Nambu-Jona-Lasinio models [13-15]
based on four-fermion interactions.

Recently a new analytical method for studying nonperturbative effects of QCD
appeared [16-20] called Method of Field Correlators (MFC) which is the generalization
of the QCD sum rule approach [4]. MFC is based on QCD and takes into account
both gluonic strings and CSB. The fundamental variables of MFC are the gluonic
condensate and the correlation length of gluonic fields. This method is based on
first principles of QCD and is supported by experiment and lattice data. It can
solve problems which can not be solved in the framework of relativistic quark
models. Such problems include: (i) the Regge slopes of meson and baryon trajectories
and their intercepts, (ii) constituent masses of quarks, (iii) radiuses and
electromagnetic polarizabilities of hadrons, (iiii) high energy scattering, and
others.

In the framework of MFC both non-perturbative effects of strong interactions
(confinement and CSB) can be explained by introducing stochastic gluon vacuum
fields with the definite fundamental correlators [16,17]. Then the linear potential
between quarks appears and it provides the confinement of quarks. The confinement of
quarks does not allow them to be observed, i.e. quarks can not move outside of hadrons
at large distances relative each other. This was confirmed by Monte-Carlo simulations and
experiments.

The Regge trajectories are asymptotically linear in this approach with a universal
slope [17]. There are different stochastic vacuum configurations which are responsible
for CSB: instantons (antiinstantons), pieces of (anti-)self-dual fields (for example
torons or randomly distributed lumps of field) and others. The necessary
requirement is to have zero fermion modes. The condensation of zero modes
leads to CSB.

So the MFC in nonperturbative QCD and the dynamics of zero
modes give the explication of the double nature of light pseudoscalar
particles (pions, kaons and others) as Nambu-Goldstone particles and as the
quark-antiquark system with a confining linear potential. It should be
noticed that confinement prevents the delocalization of zero modes over the
whole volume [18] i.e. stabilizes the phenomena of CSB.
The familiar PCAC (partial conservation of axial vector current) theorems and
the soft pion technique are reproduced in this approach [19].

There are some difficulties in evaluating meson characteristics in the general
case of the complicated string configuration. Naturally, as a first step, we make some
approximations and model assumptions to simplify the calculations. So
here we consider the straight-line string as a simple configuration and
quarks attached to the ends of the string. Such configurations were also
studied in [21]. The quark-diquark structure of baryons is employed here.

Spin degrees
of freedom are treated as a perturbation and therefore it is questionable to apply
this scheme to pions and kaons. For example $\rho $ and $K^{*}$ mesons can be considered
here because the energy shift for them due to the hyperfine spin interaction
is below $100~MeV$ [2]. Here short-range spin-orbital ${\bf L}\cdot {\bf S}$
and spin-spin ${\bf S_1}\cdot {\bf S_2}$ interactions are not taken into
account. The Coulomb like short-range contribution due to
the asymptotic freedom of QCD can be easily included.
We imply also that in the presence of light quarks the
structure of vacuum yields an area law of the Wilson loop integral.
In the present review we investigate hadrons in external,
constant, and uniform electromagnetic fields and use the path integral approach.

It is important to calculate different intrinsic characteristics of hadrons
on the basis of QCD string theory and to compare them with the
experimental values. It will be the test of this scheme. The mean radiuses
(and electromagnetic form-factors) and electromagnetic
polarizabilities of hadrons are fundamental constants which characterize the
complex structure of particles. These values for some hadrons are known from
the experimental data and therefore the estimation of them is reasonable.

The electromagnetic polarizabilities of hadrons $\alpha $, $\beta $ enter
the induced electric ${\bf D}=\alpha {\bf E}$ and magnetic ${\bf M} =\beta
{\bf H}$ dipole moments, where ${\bf E}$, ${\bf H}$ are the strengths of
electromagnetic fields. As a result there is a contribution to the
polarization potential [22] as follows $U(\alpha ,\beta )=
-(1/2)\alpha E^2-(1/2)\beta H^2$.
Electromagnetic polarizabilities are fundamental low-energy characteristics
of strong hadron interactions and therefore they can be calculated in the
framework of nonperturbative quantum chromodynamics - QCD string
theory.

The review is organized as follows. In Section 2 after describing the general
background we derive the Green function of quark-antiquark system (meson).
The effective action and Hamiltonian for mesons in external electromagnetic fields
is found in Section 3. In Section 4 we estimate the mean size of mesons on the basis of
the virial theorem. Section 5 contains the derivation of the Green Function of baryons
(three-quark system). The effective action for baryons in external electromagnetic
fields based on the proper time method and Feynman path-integrals is derived in
Section 6. Section 7 contains the calculation of average distances between quarks in
nucleons. The electric polarizabilities of nucleons are evaluated in Section 8.
We derive diamagnetic polarizabilities of protons and neutrons using the perturbative
expansion in the small magnetic fields in Section 9. In section 10 the method of
field correlators and cluster expansion are briefly described. The introduction to
the hadron electromagnetic polarizabilities is given by Appendix. In the conclusion we
made the comparison of our results with other approaches.

Units are chosen such that $\hbar =c=1$.\\

\section{Green's Function of Quark-Antiquark System}

Here we derive the Green function of quark-antiquark system using the Schwinger
proper time method and the Feynman path-integrals. Our goal is to calculate
some electromagnetic characteristics of hadrons. For this purpose we need
the effective actions for mesons and baryons in external electromagnetic fields.
The method of the Green functions will be adopted.

Let us consider the Lorentz and gauge invariant combination of two quark
colorless system (meson) in Minkowski space [17]
\begin{equation}
X_M(x,\bar{x},C)=\bar{q}_a(x)\Gamma_A\Phi_{ab}(x,\bar{x})q_b(\bar{x}),  \label{1}
\end{equation}
where $q_b(\bar{x})$ and $\bar{q}_a(x)$ are quark bispinors, $\bar{q}_a(x)=
q^{+}_a(x)\gamma _4$;
 $q^{+}_a(x)$ is the Hermitian-conjugate quark field; $\Gamma_A=1$, $\gamma_5$,
$\gamma_\mu$, $i\gamma_5\gamma_\mu$, $\Sigma _{\mu \nu }=-(i/4)[\gamma _\mu ,\gamma _\nu ]$
($\Gamma_A^{+}=\Gamma_A$, i.e. $\Gamma_A$ are Hermitian matrices);
$\gamma_\mu$ are the Dirac matrices; $a$, $b$ are colour indexes
and as usual there is a summation on repeating indexes.
We imply that quark fields $\bar{q}(x)$, $q(\bar{x})$ possess the definite flavours which
form mesons. The gauge invariance is guaranteed here by introducing the parallel
transporter [17]:
\begin{equation}
\Phi (x,\bar{x})=P\exp \left\{ ig\int_{\bar{x}}^{x}A_\mu dz_\mu \right\},
\label{2}
\end{equation}
where $P$ is the ordering operator along the contour $C$ of integration,
$g$ is the coupling constant, $A_\mu =A_\mu ^a\lambda ^a$; $A_\mu ^a$ are
the gluonic fields; $\lambda ^a$ are the Gell-Mann matrices. Although we have an
attractive feature - gauge invariance, the function (1) depends on the form
of the contour $C$.
As $X_M(x,\bar{x},C)$ is a gauge invariant object, it obeys the Gauss law on the
spacelike surface $\Sigma $. The path of integration in Eq. (2) is arbitrary.

Let the four-points $x$, $\bar{x}$, and $y$, $\bar{y}$,
be the initial and final positions of quark-antiquark, respectively. Two
particle quantum Green function is defined as [17]:
\begin{equation}
G(x,\bar{x};y,\bar{y})=\langle X_M(x,\bar{x},C)X^{+}_M(y,\bar{y},C^{\prime})\rangle,
\label{3}
\end{equation}
were $X^{+}_M(y,\bar{y},C^{\prime})$ corresponds to the final state of a meson:
\begin{equation}
X_M^{+}(y,\bar{y},C^{\prime})=\bar{q}_a(\bar{y})\Phi_{ab}(\bar{y},y)
\bar{\Gamma}_A q_b(y),  \label{4}
\end{equation}
where $\bar{\Gamma}_A=\gamma_4 \Gamma_A\gamma_4 $.
The brackets $\langle...\rangle$ mean the path-integrating over gluonic and
quark fields:
\begin{equation}
\langle X_MX^{+}_M\rangle =\int D\bar{q}DqDA_\mu \exp \left\{
iS_{QCD}\right\} X_MX^{+}_M,  \label{(5)}
\end{equation}
with the $QCD$ action $S_{QCD}$. We imply that the measure $DA_\mu $ in the
path integral (5) includes the well known gauge-fixing and Faddeev-Popov terms
for gluonic fields [23].
The Minkowski space is used here but it is not difficult to go into
Euclidean space to have well defined path-integrals.

Let us introduce the generating functional for Green's function to calculate
the path-integral (5) with respect to quark fields:
\begin{equation}
Z[\bar{\eta },\eta ]=\int D\bar{q}Dq \exp \left\{
iS_{QCD}+i\int dx\left( \bar{q}_a(x)\eta _a(x)+\bar{\eta }
_a(x)q_a(x)\right) \right\},
 \label{(6)}
\end{equation}
where we introduce the external colour anticommutative sources (Schwinger sources)
$\eta _a$, $\bar{\eta}_a$. Then the Green function (3) can be written as
\[
G(x,\bar{x};y,\bar{y}) =\int DA_\mu \biggl [\frac{\delta}{\delta \eta_{a}(x)}
\Gamma_A \Phi _{ab}(x,\bar{x})\frac{\delta}{\delta \bar{\eta}_{b}(\bar{x})}
\]
\begin{equation}
\times \frac{\delta}{\delta \eta_{m}(\bar{y})}\Phi _{mn}(\bar{y},y)
\bar{\Gamma}_A \frac{\delta }{\delta \bar{\eta }_{n}(y)}
Z[\bar{\eta},\eta ]\biggr ]_{\eta =\bar{\eta}=0}.
\label{(7)}
\end{equation}
Now it is possible to integrate the path-integral in Eq.(7) over quark
fields $\bar{q}$, $q$ as expression (6) is a Gaussian integral. We may
represent the $QCD$ action in the form of
\begin{equation}
S_{QCD}=S(A)-\int dx\bar{q}(x)\left( \gamma _\mu D_\mu +m\right) q(x),
\label{(8)}
\end{equation}
where $S(A)$ is an action for gluonic fields with the included ghost fields,
$D_\mu =\partial _\mu -iQA_\mu^{el}-igA_\mu $; $A_\mu^{el}$ is the vector
potential of an electromagnetic field, $Q$ is the charge matrix of quarks,
$Q=diag(e_1, e_2, ...e_{N_f})$, $e_i$ are charges of quarks, $N_f$ is the number of
flavours; $m$ is the quark mass matrix, $m=diag(m_1,m_2,...m_{N_f})$ and we imply
the summation on colour and flavour indexes. Thus we have just introduced the
interaction of quarks with an electromagnetic field. Inserting Eq. (8) into
Eq. (6) and integrating with respect to quark fields, we arrive at the
expression
\begin{equation}
Z[\bar{\eta },\eta ]=\det (-\gamma _\mu D_\mu -m)\exp
\left\{ iS(A)+i\int dxdy\bar{\eta }(x)S(x,y)\eta (y)\right\},
\label{(9)}
\end{equation}
where the classical quark Green function $S(x,y)$ is the solution of the
equation
\begin{equation}
\left( \gamma _\mu D_\mu +m\right) S(x,y)=\delta (x-y).  \label{(10)}
\end{equation}

Using Eq. (9) and calculating the variation derivatives in Eq. (7) we
find the quantum Green function of a meson:
\[
G(x,\bar{x};y,\bar{y})=\int DA_\mu \det (-\gamma _\mu D_\mu -m)
\exp \left\{iS(A)\right\}
\]
\[
\times \biggl ( tr \biggl [ \bar{\Gamma}_A S_{na}(y,x)
\Gamma_A \Phi_{ab}(x,\bar x)S_{bm}(\bar x,\bar y)
\Phi_{mn}(\bar y,y)\biggr ]
\]
\begin{equation}
-tr\biggl [\Phi_{mn}(\bar y,y)\bar{\Gamma}_AS_{nm}(y,\bar y)\biggr ]
tr\biggl [\Gamma_A \Phi_{ab}(x,\bar x)S_{ba}(\bar x,x) \biggr ]\biggr ).
\label{(11)}
\end{equation}
The second term in (11) corresponds to the annihilation graphs and it
does not contribute to the nonsinglet channel.
The functional determinant in Eq. (9) describes the contribution from the
vacuum polarization and gives the additional quark loops. In the quenched
approximation the fermion determinant is taken to be unity.
Neglecting quark-antiquark vacuum loops (quenched approximation) and
omitting the annihilation graph, the Green function of
a meson (the quark-antiquark system) takes the form (see also [17])
\begin{equation}
G(x,\bar x;y,\bar y)=\langle tr [ \bar{\Gamma}_A S_{na}(y,x)
\Gamma_A \Phi_{ac}(x,\bar x)S_{cm}(\bar x,\bar y)
\Phi_{mn}(\bar y,y)]\rangle_A,
\label{12}
\end{equation}
were the brackets $\langle ...\rangle_A $ mean the averaging over
the external vacuum gluonic fields with the standard measure $\exp \left[
iS\left( A\right) \right]$.

Let us derive the classical one-quark Green function which is the solution
of Eq. (10) using the Fock-Schwinger method. Starting with the approach [24],
the solution to Eq. (10) for the Green function of
a quark (in Minkowski space) is given by
\[
S(x,y)=\langle x\mid ( \widehat{D}+m_1)^{-1}\mid y
\rangle =\langle x\mid ( m_1-\widehat{D})(
m_1^2- \widehat{D}^2)^{-1}\mid y\rangle
\]
\begin{equation}
=\langle q(x) \bar{q}(y)\rangle, \label{13}
\end{equation}
where $e_1$ and $m_1$ are the charge and current mass of the quark, $\widehat{D}
=\gamma _\mu D_\mu $, $D_\mu =\partial _\mu -ie_1A_\mu ^{el}-igA_\mu $,
$A_\mu =A_\mu ^a\lambda ^a$; $\gamma _\mu $, $\lambda ^a$ are the Dirac and
Gell-Mann matrices, respectively; $A_\mu ^{el}$, $A_\mu ^a$ are the
electromagnetic and gluonic vector potentials. The inverse operator $\left(
m_1^2-\widehat{D}^2\right) ^{-1}$ can be represented in the proper time $s$
[24]:
\begin{equation}
\left( m_1^2-\widehat{D}^2\right) ^{-1}=i\int_0^\infty ds\exp \left\{
-is\left( m_1^2-\widehat{D}^2\right) \right\}. \label{14}
\end{equation}
Using the properties of the Dirac matrices $\left\{ \gamma _\mu ,\gamma _\nu
\right\} =2\delta _{\mu \nu }$ we find the squared operator
\begin{equation}
\widehat{D}^2=D_\mu ^2+\Sigma _{\mu \nu }\left( e_1F_{\mu \nu }^{el}+gF_{\mu
\nu }\right), \label{15}
\end{equation}
where
\[
\Sigma _{\mu \nu }=-\frac i4[\gamma _\mu ,\gamma _\nu ],\hspace{0.5in}F_{\mu
\nu }^{el}=\partial _\mu A_\nu ^{el}-\partial _\nu A_\mu ^{el},
\]
\begin{equation}
F_{\mu \nu }=\partial _\mu A_\nu -\partial _\nu A_\mu -ig[A_\mu ,A_\nu ],
\label{16}
\end{equation}
$\Sigma _{\mu \nu }$ are the spin matrices, and $F_{\mu \nu }^{el}$ and $F_{\mu \nu
}$ are the strength of electromagnetic and gluonic fields, respectively.
Inserting relationship (14) into Eq. (13) with the help of Eq. (15) we get
\[
S(x,y)=i\int_0^\infty ds\langle x\mid ( m_1-\widehat{D}) \]
\begin{equation}
\times\exp\{ -is [ m_1^2-D_\mu ^2-\Sigma _{\mu \nu }( e_1F_{\mu \nu
}^{el}+gF_{\mu \nu } ) ] \} \mid y \rangle.
\label{17}
\end{equation}
The exponent in Eq. (17) plays the role of the evolution operator which defines
the dynamics with the ``Hamiltonian'' $m_1^2-D_\mu ^2-\Sigma _{\mu \nu
}\left( e_1F_{\mu \nu }^{el}+gF_{\mu \nu }\right) $ with initial $\mid
y\rangle $ and final $\langle x\mid $ states where $s$ means the proper
time. Therefore it is convenient to represent the matrix element in Eq. (17) as
a path integral [25]:
\[
S(x,y)=i\int_0^\infty dsN\int_{z(0)=y}^{z(s)=x}DpDzP\left( m_1-\widehat{D}
\right) \exp \biggl [i\int_0^sdt\biggl [p_\mu \dot z_\mu -m_1^2
\]
\begin{equation}
-\left( p_\mu -e_1A_\mu ^{el}-gA_\mu \right) ^2+\Sigma _{\mu \nu }\left(
e_1F_{\mu \nu }^{el}+gF_{\mu \nu }\right) \biggr ]
\biggr ], \label{18}
\end{equation}
where $z_\mu \left( t\right) $ is the path of a quark with the boundary
conditions $z(0)=y$, $z(s)=x$, $\widehat{D}=i\gamma _\mu \left( p_\mu
-e_1A_\mu ^{el}-gA_\mu \right) $ and $P$ means ordering; $N$ is a
constant which is connected to the measure definition and it will be
chosen later. The path integration over the momenta can be rewritten in the
form (see [17,26]
\[
N\int Dp\left( m_1-\widehat{D}\right) \exp \left\{ i\int_0^sdt\left[ p_\mu
\dot z_\mu -\left( p_\mu -e_1A_\mu ^{el}-gA_\mu \right) ^2\right] \right\}
\]
\[
=N\int Dp\exp \left[ i\int_0^sdt\left( p_\mu \dot z_\mu \right) \right]
\left( m_1+\frac 12\gamma _\mu \frac \delta {\delta p_\mu }\right)
\]
\[
 \times\exp
\left\{ -i\int_0^sdt\left( p_\mu -e_1A_\mu ^{el}-gA_\mu \right) ^2\right\}
\]
\[
=N\int Dp\left( m_1-\frac i2\gamma _\mu \dot z_\mu \right) \exp \left\{
i\int_0^sdt\left[ p_\mu \dot z_\mu -\left( p_\mu -e_1A_\mu ^{el}-gA_\mu
\right) ^2\right] \right\}
\]
\begin{equation}
=N\int Dp\left( m_1-\frac i2\gamma _\mu \dot z_\mu \right) \exp \left\{
i\int_0^sdt\left[ -p_\mu ^2+\frac 14\dot z_\mu ^2+\left( e_1A_\mu
^{el}+gA_\mu \right) \dot z_\mu \right] \right\}.
\label{19}
\end{equation}
In Eq. (19) we used the integration by parts (see [26]) and made a
continuity of shifts $p_\mu \rightarrow p_\mu +e_1A_\mu ^{el}+gA_\mu $ and
then $p_\mu \rightarrow p_\mu +\dot z_\mu /2$. The constant $N$ in Eq. (19) is
defined by the relation
\begin{equation}
N\int Dp\exp \left\{ -i\int_0^sdt\left( p_\mu ^2\right) \right\} =1.
\label{20}
\end{equation}
Taking into account Eqs. (19), (20) we find from Eq. (18) the Green function of the
quark:
\[
S(x,y)=i\int_0^\infty ds\int_{z(0)=y}^{z(s)=x}Dz\;\left( m_1-\frac i2\gamma
_\mu \dot z_\mu \left( t\right) \right) P\exp {}{}{}\biggl \{i\int_0^sdt
\biggl [\frac 14\dot z_\mu ^2\left( t\right) -m_1^2
\]
\begin{equation}
+\left( e_1A_\mu ^{el}+gA_\mu \right) \dot z_\mu \left( t\right) +\Sigma
_{\mu \nu }\left( e_1F_{\mu \nu }^{el}+gF_{\mu \nu }\right) \biggr ]
\biggr \}. \label{21}
\end{equation}

\section{Effective Action for Mesons}

To get the effective action for mesons in the external electromagnetic
fields we use the Green function (21) of the quark possessing spin in the
Minkowski space which can be represented as
\[
S(x,y)=i\int_0^\infty ds\int_{z(0)=y}^{z(s)=x}Dz\;\left( m_1-\frac i2\gamma
_\mu \dot z_\mu \left( t\right) \right) P\exp \biggl \{i\int_0^s \biggl [
\frac 14\dot z_\mu ^2(t)-m_1^2
\]
\begin{equation}
+e_1\dot z_\mu (t)A_\mu ^{el}(z)+\Sigma _{\mu \nu }\left( e_1F_{\mu \nu
}^{el}+gF_{\mu \nu} \right) \biggr ]dt\biggr \}\Phi (x,y), \label{22}
\end{equation}
where $\Phi (x,y)$
is the path ordered product (2).
Inserting Eq. (22) for a quark ($S(x,y$)) and antiquark ($S(\bar y,\bar x$))
Green functions into Eq. (12) we find the
expression for the Green function of a meson:
\[
G(x,\bar x;y,\bar y)=tr\int_0^\infty ds\int_0^\infty d\bar
s\int_{z(0)=x}^{z(s)=y}Dz\bar{\Gamma}_{A}\left( m_1-\frac i2\gamma _\mu
\dot z_\mu \left( t\right) \right)
\]
\[
\times \langle P_\Sigma P_A \exp \left\{ ig\Sigma _{\mu \nu } \int_0^s
F_{\mu \nu}\left( z\right) dt\right\}\Gamma_A \exp\left\{-ig\Sigma _{\mu \nu }
\int_0^{\bar s}F_{\mu \nu }\left( \bar z\right) d\bar
t\right\} W(C)\rangle_{A}
\]
\[
\times\int_{\bar z(0)=\bar y}^{\bar z(\bar s)=\bar x}D\bar z\left( m_2-\frac
i2\gamma _\mu \dot z_\mu \left( \bar t\right)
\right)
\]
\[
\times \exp \biggl \{i\int_0^s\biggl [\frac 14\dot z_\mu
^2(t)-m_1^2+e_1\dot z_\mu (t)A_\mu ^{el}(z)
+e_1\Sigma _{\mu \nu }F_{\mu \nu }^{el}\left( z\right) \biggr ]dt
\]
\begin{equation}
+i\int_0^{\bar s}\left[ \frac 14 \dot z_\mu ^2(\bar
t)-m_2^2+e_2\dot z_\mu (\bar t)A_\mu ^{el}(\bar
z)+e_2\Sigma _{\mu \nu }F_{\mu \nu }^{el}\left( \bar z\right) \right] d\bar
t \biggr \},
\label{23}
\end{equation}
where $P_{\Sigma}$ and $P_A$ are the ordering operators of the spin matrices
$\Sigma_{\mu \nu }$ and gluonic fields, respectively; $e_1$, $e_2$ are
charges and $m_1$, $m_2$ are current masses
of the quark and antiquark; $z_\mu (t)$, $\bar z_\mu (\bar t)$ are the paths
of the quark and antiquark with the boundary conditions $z_\mu (0)=x_\mu $,
$z_\mu (s)=y_\mu $, $\bar z_\mu (0)=\bar y_\mu $, $\bar z_\mu (\bar s)=\bar
x_\mu $ and $\dot z_\mu (t)=\partial z_\mu /\partial t$.
As compared with [17, 27]
we added the interaction of charged quarks with the external electromagnetic
fields. The gauge - and Lorenz - invariant Wilson loop operator is given by
\begin{equation}
W(C)=\frac{1}{N_C}trP\exp\left\{ ig\int_CA_\mu dz_\mu \right\}, \label{24}
\end{equation}
where $N_C$ is the color number, and $C$ is the closed contour of lines $x\bar x$
and $y\bar y$ connected by paths $z(t)$, $\bar z(\bar t)$ of the quark and
antiquark. The Wilson operator (24) contains both the perturbative and
nonperturbative interactions between quarks via gluonic fields $A_\mu $ with
the QCD coupling constant $g$, and it is the amplitude of the process of
creation and annihilation of quarks and antiquarks.

Using non-Abelian Stokes theorem [28, 17], the Wilson loop is rewritten as
\begin{equation}
W(C)=\frac{1}{N_C}trP\exp\left\{ ig\int_{\Sigma}F_{\mu\nu}(z,z_0)
d\sigma_{\mu\nu}(z) \right \}, \label{25}
\end{equation}
where $\Sigma$ is an arbitrary surface bounded by the contour $C$, and
$F_{\mu\nu}(z,z_0)$ is given by
\begin{equation}
 F_{\mu\nu}(z,z_0)=\Phi(z_0,z)F_{\mu\nu}(z)\Phi(z,z_0).
\label{26}
\end{equation}
The coordinate $z_0$ and path of the transporter $\Phi(z,z_0)$
belongs to the surface $\Sigma$, and they are arbitrary [17]. The contour
$C$ can be parametrized by the four vector $z_{\mu}(\xi)$, where two
dimensional coordinate $\xi=(\xi_1,\xi_2)$ induces the metric tensor
of the surface $\Sigma$: $g_{\mu\nu}^{ab}(\xi)=[\partial_a z_\mu (\xi)]
[\partial_b z_\nu (\xi)]$ $(\partial_a =\partial/\partial\xi_a)$.
The infinitesimal element of the surface $\Sigma$ is given by
\begin{equation}
 d\sigma_{\mu\nu}(z)=\epsilon^{ab}g_{\mu\nu}^{ab}(\xi)d^2 \xi,
\label{27}
\end{equation}
where $\epsilon^{ab}=-\epsilon^{ba}, \epsilon^{12}=1$.
In accordance with the approach in [17], spin interactions can be treated as
perturbations, e.g. for $\rho$ and $K^*$ mesons but not for Nambu-Goldstone
$\pi$- and $K$-mesons. To construct the
expressions in spin interactions we write the relationship [17]
\[
\langle \exp \left\{ ig\Sigma _{\mu \nu }\left[ \int_0^sF_{\mu \nu
}\left( z\right) dt\right ] \right\} W(C)\rangle_A
\]
\begin{equation}
= \exp \left\{ \Sigma _{\mu \nu }\left[ \int_0^sdt\frac {\delta}{\delta \sigma
_{\mu \nu }\left( t\right)}\right] \right\}
\langle W(C) \rangle_A,
 \label{28}
\end{equation}
where $\delta \sigma _{\mu \nu }\left( t\right) $ is the surface around the
point $z_\mu \left( t\right) $. The zeroth order in spin-orbit and spin-spin
interactions corresponds to neglecting the terms $e_i \Sigma_{\mu \nu }
F_{\mu\nu}^{el}$, $g\Sigma _{\mu \nu }F_{\mu\nu }$
in (23).

To receive the effective action for mesons from Eq. (23) one needs to estimate
the Wilson loop integral (24) which carries very important information about
an interaction of quarks.

Let us consider the case when the distance
between quarks is greater than the time fluctuations of the gluonic fields
($r\gg T_g$). The Monte-Carlo calculations [29] gave $
T_g\simeq 0.2\div 0.3fm$. We imply that the characteristic quark relative
distance is $r\simeq 1~fm$. This assumption is confirmed by the lattice data [30]
and by the calculation of the quark-antiquark relative coordinate [31].

The average Wilson integral (24) at large distances in accordance with the
area law can be represented in the Minkowski space as
\begin{equation}
\langle W(C) \rangle_A = \exp(i\sigma S_{min}), \label{29}
\end{equation}
where $\sigma $ is the string tension and $S_{min}$ is the area of the minimal
surface inside of the contour $C$. This behavior of the Wilson loop integral
is the consequence of the nonperturbative interactions of a quark and antiquark,
and corresponds to the linear potential. The area law of the Wilson loop is the
criterion of confinement [32]. In this case the gluonic field between a quark and
antiquark forms a string, and the string tension $\sigma $ is the energy of a string
per unit length. In accordance with the lattice data (see e.g. [17]) $\sigma
\simeq 0.2 ~GeV^2$.
It is impossible to calculate a dimensional parameter $\sigma $ using the perturbative
expansion in the small coupling constant $g$. There is a dimensional parameter of
QCD in the high-energy limit - the scale parameter $\Lambda_{QCD}\simeq 200 ~MeV$:
\begin{equation}
 \Lambda_{QCD}^2=\mu^2\exp\left( -\frac{16\pi^2}
{((11/3)N_C -(2/3)N_f)g^2(\mu^2)}\right ),
\label{30}
\end{equation}
where $\mu$ is a normalization mass.
According to the phenomenon of dimensional transmutation, dimensional parameters are
proportional to some power of $\Lambda_{QCD}$. It is easy to see from Eq. (30) that
all coefficients of the expansion of $\Lambda_{QCD}$ in powers of $g^2$ are zero.
Therefore, $\sigma$ is proportional to $\Lambda_{QCD}^2$ and the QCD strings are
described by the nonperturbative approach.

In the case of perturbative interactions at small distances (high energy), the Coulomb
potential dominates, which does not confine quarks. QCD is an asymptotically free
theory and at high energy the coupling constant is a small parameter. In this regime
the gauge field is not concentrated in the tube (string) but fills the whole space.
Then the asymptotic form of the Wilson loop (for a smooth contour $C$) for this case
is given by the perimeter law:
\begin{equation}
\langle W(C) \rangle_A =\exp\{i\gamma L(C)\}\langle W(C)_{ren}\rangle_A ,
\label{31}
\end{equation}
where the appearance of the finite value $\langle W(C)_{ren}\rangle $ is due to
the renormalization; $\gamma$
is a dimensional constant, and $L(C)$ is the length of the contour ($L(C) =\int_0^1 ds
\sqrt{\dot{z}_\mu^2(s)}$, $z_\mu(0)=z_\mu(1)$). When the distance between quarks
$r<0.25 ~fm$, the system is in the Coulomb phase and we have the perimeter law
behavior of the Wilson averaged integral (31). The confining phase (nonperturbative)
occurs at large distances and gives the area law (29). For wide class of contours
without self-intersections, the Wilson loop integral is multiplicatively renormalizable
[33] and is given by
\begin{equation}
\langle W(C) \rangle_A =\langle W(C)\rangle_{pert}\langle W(C)\rangle_{nonpert},
\label{32}
\end{equation}
where the first factor in right-hand side of Eq. (32) describes the perturbative contribution
in the Wilson integral (31), and the second describes confinement.

The surface $S_{min}$ in Eq. (29) can be parametrized by the Nambu-Goto form [8]
\begin{equation}
S_{min}=\int_0^Td\tau \int_0^1d\beta \sqrt{(\dot w_\mu w_\mu ^{\prime })^2-\dot
w_\mu ^2w^{\prime }{}_\nu ^2}, \label{33}
\end{equation}
where $\dot w_\mu =\partial w_\mu /\partial \tau $, $w_\mu ^{\prime
}=\partial w_\mu /\partial \beta $.

Using the approximation [17] that the
coordinates of the string world surface $w_\mu (\tau ,\beta )$ can be taken
by straight lines for the minimal surface we write
\begin{equation}
w_\mu (\tau ,\beta )=z_\mu (\tau )\beta +\bar z_\mu (\tau )(1-\beta ),
\label{34}
\end{equation}
where $\tau $ is implied to be the proper time parameter for both
trajectories $\tau =(tT)/s=(\bar tT)/\bar s$. Thus we ignore vibrations of the
string.

For uniform static
external electromagnetic fields we have the representation of the vector
potential through the strength tensor $F_{\mu \nu }^{el}$:
\begin{equation}
A_\nu ^{cl}(z)=\frac 12F_{\mu \nu }^{el}z_\mu ,\qquad A_\nu ^{cl}(\bar z)=\frac
12F_{\mu \nu }^{el}\bar z_\mu. \label{35}
\end{equation}
The paths $z_\mu $, $\bar z_\mu $ are expressed via the center of mass
coordinate $R_\mu $ and the relative coordinate $r_\mu $ [17],
\begin{equation}
\bar z_\mu (\tau)=R_\mu -\frac{\bar s}{s+\bar s}r_\mu ,\qquad z_\mu (\tau
)=R_\mu +\frac s{s+\bar s}r_\mu \label{36}
\end{equation}
with the boundary conditions for $R_\mu (\tau )$, $r_\mu (\tau )$:
\[
R_\mu (0)=\frac{\mu _1y_\mu +\mu _2\bar y_\mu }{\mu _1+\mu _2},~~R_\mu (T)=
\frac{\mu _1x_\mu +\mu _2\bar x_\mu }{\mu _1+\mu _2},
\]
\begin{equation}
r_\mu (0)=y_\mu
-\bar y_\mu ,~~r_\mu (T)=x_\mu -\bar x_\mu. \label{37}
\end{equation}
The integration with respect to $z_\mu $, $\bar z_\mu $ in (23) is replaced
by the integration over new variables $R_\mu $, $r_\mu $. As $\tau $ is a
common time for the quark and antiquark (the time of the meson) the
parametrization $z_\mu =({\bf z},i\tau )$, $\bar{z}_\mu =({\bf \bar z},i\tau)$
is possible [17]. This leads to the constraints: $R_0(\tau )=\tau $,
$r_0(\tau )=0$. In accordance with the approach [17] we introduce the
dynamical masses $\mu _1$, $\mu _2$ by relationships
\begin{equation}
\mu _1=\frac T{2s},\qquad \mu _2=\frac T{2\bar s}. \label{38}
\end{equation}
Replacing the integration with respect to $s$, $\bar s$ in (23) by the
integration over $d\mu _1$ and $d\mu _2$ with the help of Eqs. (29), (33) -
(36) we find [31] the two-point function in the zeroth order in spin interactions
\begin{equation}
G(x,\bar x;y,\bar y)=T^2\int_0^\infty \frac{d\mu _1}{2\mu _1^2}
\int_0^\infty \frac{d\mu _2}{2\mu _2^2}\int DRDr\; exp\left\{
iS_{eff}\right\} \label{39}
\end{equation}
with the effective action
\[
S_{eff}=\int_0^Td\tau \biggl [-\frac{m_1^2}{2\mu _1}-\frac{m_2^2}{2\mu _2}
+\frac 12\left( \mu _1+\mu _2\right) \dot R_\mu ^2+\frac 12\tilde \mu \dot
r_\mu ^2
\]
\[
+\frac 12F_{\nu \mu }^{el}e\left( \dot R_\mu R_\nu +\frac 14\dot r_\mu r_\nu
\right) -\frac q4F_{\nu \mu }^{el}\left( \dot R_\mu r_\nu
+\dot r_\mu R_\nu\right)
\]
\begin{equation}
-\int_0^1d\beta \sigma _0\sqrt{(\dot w_\mu w_\mu ^{\prime })^2-\dot
w_\mu ^2w^{\prime }{}_\nu ^2}\biggr ],
 \label{40}
\end{equation}
where $w_\mu =R_\mu +[\beta -\mu _1/(\mu _1+\mu _2)]r_\mu, \tilde \mu =\mu
_1\mu _2/(\mu _1+\mu _2)$ is the reduced mass of the quark-antiquark system,
$e=e_1+e_2$, $q=e_1-e_2$. As a first step we are interested here in the
spinless part and therefore the preexponential terms $\left( m_1-\frac
i2\gamma _\mu \dot z_\mu \left( t\right) \right) $, $\left( m_2-\frac
i2\gamma _\mu \stackrel{\stackrel{.}{\_}}{z}_\mu \left( \bar t\right)
\right) $ and the constant matrices $\ \Sigma _{\mu \nu }F_{\mu \nu }^{el}$
were omitted in Eq. (39). The expression (40) defines the effective
Lagrangian for mesons in external uniform static electromagnetic
fields in accordance with the formula $S_{eff}=\int_0^T{\cal L} _{eff}d\tau
$. The expression (40) looks like nonrelativistic one at $F_{\mu \nu }=0$,
but it is not. The author of [17] showed that the relativism is
contained here due to the $\tilde \mu $ dependence and the spectrum is
similar to that of the relativistic quark model.

The mass of the lowest states can be found on the basis of the relationship
[34]
\[
\int DRDr\exp\left\{ iS_{eff}\right\} =
\]
\begin{equation}
\langle R=\frac{\mu_1x+\mu_2\bar x}{\mu_1+\mu_2},r=x-\bar x
\mid\exp\{-iT{\cal M}(\mu_1,\mu_2)\}\mid
R=\frac{\mu _1y+\mu _2\bar y}{\mu_1+\mu _2} ,r=y-\bar y \rangle, \label{41}
\end{equation}
where the mass ${\cal M}(\mu _1,\mu _2)$ is the eigenfunction of the
Hamiltonian. After that the Green function (39  ) is derived by integrating
(40) over the dynamical masses $\mu _1$, $\mu _2$. In accordance with [17]
we estimate the last integration on $d\mu _1$, $d\mu _2$ using the steepest
descent method which gives a good accuracy when the Minkowski time $
T\rightarrow \infty $. To have the correct formulas, it is necessary to go
into Euclidean space and return into Minkowski space on completing the
functional integration. We use this procedure.

The last term in (40) can be represented by the relation
\[
\sigma\int_0^1d\beta \sqrt{(\dot w_\mu w_\mu ^{\prime })^2-\dot w_\mu ^2
w^{\prime 2}_\nu }
\]
\begin{equation}
=\sigma\int_0^1d\beta \sqrt{b_0+2b_1\left( \beta -\frac{\mu _1}{ \mu_1+\mu_2}
\right)+b_2 \left(\beta -\frac{\mu_1}{\mu_1+\mu _2}\right)^2},
 \label{42}
\end{equation}
where
\[
b_0=\left(r_\mu \dot{R}_\mu \right)^2-r_\mu^2 \dot{R}_\nu^2,~~b_1=\left (r_\mu
\dot{r}_\mu\right)\left (r_\nu\dot{R}_\nu\right)-r_\mu^2\left(\dot{r}_\nu
\dot{R}_\nu\right),~~b_2=\left(r_\mu \dot{r}_\mu \right)^2-r_\mu^2 \dot{r}_\nu^2.
\]
In the pure potential regime at low orbital excitations of the string when the
orbital quantum number $l$ is small, expression (42) is equal to
$\sigma\sqrt{{\bf r}^2}$. As the equalities $R_0(\tau )=\tau $,
$r_0(\tau )=0$ are valid, the dynamical quantities are 3-dimensional.
From Eqs. (40), (42) using the standard procedure we find the
canonical three-momenta corresponding to the center of mass coordinate $R_\mu $
and the relative coordinate $r_\mu $:
\[
\Pi_k=\frac{\partial {\cal L}_{eff}}{\partial \dot {R_k}}=(\mu_1+\mu_2) \dot
{R_k}+\frac e2F_{\nu k}^{el}R_\nu +\frac q4F_{\nu k}^{el}r_\nu
\]
\[
-\sigma\int_0^1d\beta \frac{\left(r_\mu \dot{R}_\mu \right)r_k-r_\mu^2 \dot{R}_k
+\left( \beta -\frac{\mu _1}{ \mu_1+\mu_2}\right)\left[\left (r_\mu
\dot{r}_\mu\right)r_k-r_\mu^2\dot{r}_k \right]}
{\sqrt{b_0+2b_1\left( \beta -\frac{\mu _1}{ \mu_1+\mu_2}
\right)+b_2 \left(\beta -\frac{\mu_1}{\mu_1+\mu _2}\right)^2}},
\]
\[
 p=\frac{\partial {\cal L}_{eff}}{\partial \dot {r_k}}=\tilde \mu \dot
{r_k}+\frac e8F_{\nu k}^{el}r_\nu +\frac q4F_{\nu k}^{el}R_\nu
\]
\begin{equation}
-\sigma\int_0^1d\beta \frac{\left( \beta -\frac{\mu _1}{ \mu_1+\mu_2}\right)
\left[\left(r_\mu \dot{R}_\mu \right)r_k-r_\mu^2 \dot{R}_k\right]
+\left( \beta -\frac{\mu _1}{ \mu_1+\mu_2}\right)^2\left[\left (r_\mu
\dot{r}_\mu\right)r_k-r_\mu^2\dot{r}_k \right]}
{\sqrt{b_0+2b_1\left( \beta -\frac{\mu _1}{ \mu_1+\mu_2}\right)
+b_2 \left(\beta -\frac{\mu_1}{\mu_1+\mu _2}\right)^2}}, \label{43}
\end{equation}
The Hamiltonian ${\cal H}=\pi _k\dot r_k+\Pi _k\dot R_k-{\cal L}_{eff}$
found from (40) with the help of (42), (43) takes the form
\[
{\cal H}=\frac{m_1^2}{2\mu_1}+\frac{m_2^2}{2\mu_2}+\frac{\mu_1+\mu_2} {2}+
\frac{ \mu_1+\mu_2}{2}\dot {\bf R}^2+\frac{\tilde \mu }2\dot {\bf r}^2-\frac e2(
{\bf E} {\bf R})-\frac q4({\bf E}{\bf r})
\]
\begin{equation}
+\sigma\int_0^1d\beta \frac{{\bf r}^2}
{\sqrt{b_0+2b_1\left( \beta -\frac{\mu _1}{ \mu_1+\mu_2}\right)
+b_2 \left(\beta -\frac{\mu_1}{\mu_1+\mu _2}\right)^2}},
\label{44}
\end{equation}
so that the equation for the eigenvalues is given by
\begin{equation}
{\cal H}\Phi ={\cal M}(\mu_1,\mu_2)\Phi.
\label{45}
\end{equation}
 As the canonical momentum ${\bf \Pi}$ corresponding to the center of mass coordinate
 is a constant due to the conservation law, i.e. ${\bf \Pi}=const$, we can put $\dot{R}_k=0$,
$\dot{R}_4=i$. At this choice parameters occurring Eq. (44) become (at $r_4=0$) $b_0={\bf r}^2$,
$b_1=0$, $b_2=-({\bf r}\times \dot{{\bf r}})^2$. Then the last term in Eq. (44) giving the
contribution to the energy due to the string is given by
\[
{\cal H}_{string}=\sigma\int_0^1d\beta \frac{{\bf r}^2}
{\sqrt{{\bf r}^2
-({\bf r}\times \dot{{\bf r}})^2 \left(\beta -\frac{\mu_1}{\mu_1+\mu _2}\right)^2}},
\]
The expression ${\cal H}_{string}$ makes sense here when the value under the root squared is
positive. This condition  is realized at $({\bf r}\times \dot{{\bf r}})^2<2{\bf r}^2$
(if $\mu_1=\mu_2$), i.e. for low orbital quantum numbers $l$. After the integration over the
parameter $\beta$ we get
\[
{\cal H}_{string}=\frac{\sigma {\bf r}^2}{\sqrt{({\bf r}\times \dot{{\bf r}})^2 }}
\biggl [ \arcsin\left (\frac{\mu_1}{\mu_1+\mu_2}\sqrt{\frac{({\bf r}\times \dot{{\bf r}})^2 }
{{\bf r}^2}}\right)
\]
\begin{equation}
+ \arcsin\left (\frac{\mu_2}{\mu_1+\mu_2}\sqrt{\frac{({\bf r}\times
\dot{{\bf r}})^2 }{{\bf r}^2}}\right)\biggr].
\label{46}
\end{equation}
In the potential regime at low orbital quantum number when
$({\bf r}\times \dot{{\bf r}})^2\ll{\bf r}^2$   Eq. (46) is replaced by
\[
{\cal H}_{string}=\sigma_0\sqrt{{\bf r}^2}
\]
and it describes the linear potential which guarantees confinement of quarks.

The terms contained the strength of the electric field in Eq. (44) describe the
interaction of the dipole electric moment ${\bf d}$ with an external
electric field. Using the definitions we have
\begin{equation}
\frac e2({\bf E}{\bf R})+\frac q4({\bf E}{\bf r})=\frac 12(e_1 {\bf r_1}+e_2
{\bf r_2}){\bf E}={\bf d}{\bf E} \label{47}
\end{equation}
and the interaction energy of the electric dipole moment with a uniform
static electric field is $U=-{\bf d}{\bf E}$.

In the center of mass system when ${\bf R}=const$ and at $l=0$, ${\bf E}=0$,
${\bf H}=0$, taking into account Eq. (43), we find (see [17]) from Eq. (44) the
expression
\begin{equation}
{\cal H}_0=\frac{m_1^2}{2\mu_1}+\frac{m_2^2}{2\mu_2}+\frac{\mu_1+\mu_2}{2}+
\frac{{\bf p}^2}{2\tilde \mu}+\sigma\sqrt{{\bf r}^2}.
\label{48}
\end{equation}
The Hamiltonian (48) describes two quarks which are connected by the non-rotating
string. Finding extremum of ${\cal H}_0$ in $m_1$ and $m_2$: $\partial {\cal H}_0
/\partial \mu_1=0$, $\partial {\cal H}_0/\partial \mu_2=0$, one arrives at
\[
\mu_{10}=\sqrt{{\bf p}^2+m_1^2},~~\mu_{20}=\sqrt{{\bf p}^2+m_2^2},
\]
 \begin{equation}
{\cal H}_{1}=\sqrt{{\bf p}^2+m_1^2}+\sqrt{{\bf p}^2+m_2^2}+\sigma\sqrt{{\bf r}^2}.
\label{49}
\end{equation}
Thus we come to the Hamiltonian of the relativistic quark model (RQM) (see [35]).
In the approach considered $\mu_1$ and $\mu_2$ are dynamical masses and $m_1$, $m_2$ are
current masses of quarks. In RQM Eq. (49) is used for dynamical masses $m_1$, $m_2$ of
the order of $200 ~MeV$ and arbitrary orbital number $l$. So the Hamiltonian (49) ignores
the rotation of the string and gives the slope of the Regge trajectories $1/8\sigma$ [17].
Besides, in RQM to get the mass of $\rho$-meson one needs to subtract $700\div800$ MeV.
Such large negative constant introduced by hand in RQM has nonperturbative nature. The
appearance of this constant is due to selfenergies of quarks and  can be explained in
the framework of CSB in QCD string approach [17].

At large orbital quantum number $l$ the Hamiltonian becomes [17]
 \[
{\cal H}_{2}^2=2\pi\sigma\sqrt{l(l+1)},
\]
so that the Regge slope is equal to $1/2\pi\sigma$.

\section{Mean size of Mesons}

Now we find the wave function of the ground state and estimate the mean size of mesons.
The equation for the eigenfunction
$\Phi_0 $ of the auxiliary ``Hamiltonian''
\[
\tilde {{\cal H}}_0 ={\cal H}_0
-\frac {m_1^2}{2\mu _1}-\frac {m_2^2}{2\mu _2}-\frac {(\mu _1+\mu _2)}{2}
\]
is given by
\begin{equation}
\biggl (\frac 1{2\tilde \mu }{\bf p}^2+\sigma _0\sqrt{{\bf r}^2}\biggr )
\Phi_0 =\epsilon (\mu)\Phi_0,
\label{50}
\end{equation}
so that $\tilde {{\cal H}}_0\Phi_0=\epsilon (\mu)\Phi_0$, where $\epsilon (\mu)$ is
the eigenvalue.
We can apply equation (50) to the leading trajectories with light quarks with
masses $m_1=m_2 \equiv m$, $\mu_1=\mu_2 \equiv \mu$ ($\tilde {\mu} =\mu/2$)
for $\rho$ mesons and when $m_1\neq m_2$, $\mu_1\neq \mu_2$ for $K^*$ mesons [36].
In quantum theory instead of the path integration in ${\bf r}$ we can
use the replacement $p_{k} \rightarrow - i\partial/\partial r_{k}$.
Then Eq. (50) becomes
\begin{equation}
\biggl (-\frac 1{2\tilde \mu }\frac{\partial^2}{\partial r_i^2}+
\sigma _0\sqrt{{\bf r}^2}
\biggr )
\Phi_0 =\epsilon (\mu)\Phi_0, \label{51}
\end{equation}
Equation (51) gives the discrete values of the energy $\epsilon (\mu)$ due
to the shape of the potential energy. The numerical solution of
Eq. (51) was obtained in [40]. It is useful to find the solution to equation
(51) for the ground state in analytical form. After introducing the
variables $\rho _k=(2 \tilde \mu \sigma_0)^{1/3}r_k$, $\epsilon (\tilde{\mu})=
(2 \tilde{\mu} )^{-1/3}\sigma_0^{2/3}a(n)$ [17], Eq. (51) becomes
\begin{equation}
\left( -\frac{\partial ^2}{\partial \rho _i^2}+\rho \right) \Phi_0 (\rho
)=a(n)\Phi_0 (\rho ),
\label{52}
\end{equation}
where $\rho =\sqrt{\rho _1^2+\rho _2^2+\rho _3^2}$.
The solution to Eq. (52) may be chosen in the form $\Phi_0 (\rho
)=R(\rho )Y_{lm}(\theta ,\phi )$, where $Y_{lm}(\theta ,\phi )$ are
spherical functions. After setting the variable $R(\rho )=\chi (\rho)/\rho $ we
come to the equation for the radial function
\begin{equation}
\chi ^{^{\prime \prime }}(\rho )+\left( a(n)-\rho -\frac{l(l+1)}{\rho ^2}
\right) \chi (\rho )=0, \label{53}
\end{equation}
where $\chi ^{^{\prime \prime }}(\rho )=\partial ^2\chi (\rho )/\partial
\rho ^2$, and $l$ is an orbital quantum number. The solutions to Eq. (53)
for the ground state $l=0$ are the Airy functions $Ai(\rho -a(n))$, $Bi(\rho
-a(n))$ [41]. The finite solution to Eq. (53) at $\rho \rightarrow \infty $ $
(l=0)$ is
\begin{equation}
\chi (\rho )=NAi(\rho -a(n)). \label{54}
\end{equation}
The constant $N$ can be found from the normalization condition
\begin{equation}
\int_0^\infty \chi ^2(\rho )d\rho =1. \label{55}
\end{equation}
The requirement that this solution satisfies the condition $\chi
(0)=NAi(-a(n))=0$ gives the Airy function zeroes [41] $a(1)\equiv a_1=2.3381$,
 $a(2)\equiv a_2=4.0879$ and so on. The main quantum number $n=n_r+l+1$,
where $n_r$ is the radial quantum number which defines the number of zeroes
of the function $\chi (\rho )$ at $\rho >0$. For the ground state we should
take the solution (54) at $a(n)=a_1$~ (here $n_r=0$, $l=0$, $n=1$):
\begin{equation}
\chi _0(\rho )=N_0Ai(\rho -a_1). \label{56}
\end{equation}
For the excited states it is necessary to choose the corresponding
value of $n=n_r+l+1$.

Now let us estimate the mean-squared radius for the state which is described
by the function $\Phi_0$ (the solution of Eq. (50)). Multiplying Eq. (50)
by the conjugated function $\Phi_0 ^{*}$ and integrating over the volume we
find the relations
\[
\left\langle T\right\rangle +\left\langle U\right\rangle =\epsilon (\tilde
\mu ),
\]
\begin{equation}
\left\langle T\right\rangle =-\frac {1}{2\tilde \mu} \int \Phi_0 ^{*}\partial _k^2\Phi_0
dV,~~~\left\langle U\right\rangle =\sigma _0\int \sqrt{{\bf r}^2}\Phi_0
^{*}\Phi_0 dV. \label{57}
\end{equation}
It is seen from Eqs. (57) that the mean potential energy $\left\langle
U\right\rangle =\sigma _0\left\langle \sqrt{{\bf r}^2}\right\rangle $ is
connected with the mean diameter $\left\langle \sqrt{{\bf r}^2}
\right\rangle $ ~(because ${\bf r}$ is the relative coordinate and quarks
move around their center of mass), which defines the size of mesons. In
accordance with the virial theorem [42] we have the connection of the mean
kinetic energy with the mean potential energy:
\begin{equation}
2\left\langle T\right\rangle =k\left\langle U\right\rangle, \label{58}
\end{equation}
where $k$ is defined from the equality $U(\lambda r)=\lambda ^kU(r)$. In our
case of the linear potential $k=1$ and from Eqs. (57), (58) we get
\begin{equation}
\left\langle U\right\rangle =\frac 23\epsilon (\tilde \mu )=\frac 23(2\tilde
\mu )^{-1/3}\sigma _0^{2/3}a(n).
\label{59}
\end{equation}
The use of the steepest descent method for the estimation of the
integration in $\mu $~ (at ${\bf H}=0$) leads to the conditions [17]:
\begin{equation}
\frac{\partial {\cal M}(\mu _1,\mu _2)}{\partial \mu _1}=0,~~~~\frac{
\partial {\cal M}(\mu _1,\mu _2)}{\partial \mu _2}=0, \label{60}
\end{equation}
where the mass of the ground state ${\cal M}(\mu _1,\mu _2)$ is given by
[see Eqs. (44), (45)]
\begin{equation}
{\cal M}(\mu_1,\mu_2)=\frac{m_1^2}{2\mu_1}+\frac{m_2^2}{2\mu_2}+\frac{
\mu_1+\mu_2} 2+(2\tilde \mu )^{-1/3}\sigma_0^{2/3}a(n). \label{61}
\end{equation}
Here we consider the more general case as compared with [17] when $\mu _1\neq
\mu _2$~ ($m_1\neq m_2$). This case is realized for $K^*$ mesons. It is
assumed that the current mass of u,d-quarks ($m_u=5.6\pm 1.1~MeV$, $
m_d=9.9\pm 1.1~MeV$ [43]), $m_1$ is much less than the dynamical mass $\mu _1$
($\mu _1\simeq 330~MeV$), i.e. $m_1\ll \mu _1$ and the mass of s-quark $m_2$
~($m_s=199\pm 33~MeV$ [43]) is comparable with $\mu _1$ but $m_2<\mu _1$.
Using these assumptions we neglect the term $m_1^2/2\mu _1$ in Eq. (61) and
from Eqs. (60) have the equations
\begin{equation}
(2\tilde \mu \sigma _0)^{2/3}a(n)=3\mu _1^2,~~~~3m_2^2+(2\tilde \mu \sigma
_0)^{2/3}a(n)=3\mu _2^2. \label{62}
\end{equation}
From Eqs. (62) we arrive to the expression for the dynamical mass $\mu _2$ (for
s-quark):
\begin{equation}
\mu _2=\sqrt{\mu _1^2+m_2^2}. \label{63}
\end{equation}
To find $\mu _1$ the perturbation in the parameter $m_2^2/\mu _1^2$ will be
assumed. Using the relation $\mu _2\simeq \mu _1(1+m_2^2/(2\mu _1^2))$ which
is obtained from Eq. (63) and the definition of the reduced mass $\tilde \mu
=\mu _1\mu _2/(\mu _1+\mu _2)$ from Eqs. (62) we arrive at the equation
\begin{equation}
\mu _1\simeq \sqrt{\sigma _0}\left( \frac{a(n)}3\right) ^{3/4}\left( 1+\frac{
m_2^2}{8\mu _1^2}\right). \label{64}
\end{equation}
In zeroth order we come to the value $\mu _0\equiv \mu _1^{(0)}=\sqrt{
\sigma _0}(a(n)/3)^{3/4}$ [17]. The next order gives the relationship
\begin{equation}
\mu _1\simeq \sqrt{\sigma _0}\left( \frac{a(n)}3\right) ^{3/4}+\frac{m_2^2}{
8 \sqrt{\sigma _0}}\left( \frac 3{a(n)}\right) ^{3/4}. \label{65}
\end{equation}
In a particular case $m_2=0$ we arrive at $\mu _1=\mu _2=\mu _0=\sqrt{\sigma
_0}(a(n)/3)^{3/4}$ [17]. The value of the string tension $\sigma
_0=0.15~GeV^2 $ was found from a comparison of the experimental slope of
the linear Regge trajectories, $\alpha ^{\prime }=0.85~GeV^{-2}$, and the
variable $\alpha ^{\prime }=1/8\sigma _0$ [17]. It leads for the lowest state
$n_r=0,l=0,a(1)=2.3381$ to the value $\mu _0=321~MeV$ [17]. This means that
for $\rho $-mesons when $m_1=m_u$, $m_2=m_d$ we have the dynamical masses of
$u$, $d$-quarks $\mu _1=\mu _2=\mu _0$. For $K^*$-mesons using Eq. (63) and $
m_2=m_s\simeq 200~MeV$ [43] from Eq. (65) we get the reasonable values
\begin{equation}
\mu _1\simeq 337~MeV,~~\mu _2\simeq 392~MeV,~~\tilde \mu \simeq 181~MeV.
\label{66}
\end{equation}
Inserting the equation $\left\langle U\right\rangle =\left\langle \sigma _0\sqrt{
{\bf r}^2}\right\rangle $ into the left-hand side of Eq. (59) produces the expression
\begin{equation}
\left\langle \sqrt{{\bf r}^2}\right\rangle =\frac 23(2\tilde \mu \sigma
_0)^{-1/3}a(n). \label{67}
\end{equation}
From Eqs. (63), (65) using the first order in the parameter $m_2^2/\mu _1^2$
we find
\begin{equation}
2\tilde \mu \simeq \mu _0\left( 1+\frac{3m_2^2}{8\mu _0^2}\right)
~~~~~~\left( \mu _0=\sqrt{\sigma _0}\left( \frac{a(n)}3\right) ^{3/4}\right).
 \label{68}
\end{equation}
Eq. (67) with the help of Eq. (68) gives the approximate relation for the
mean relative coordinate
\begin{equation}
\left\langle \sqrt{{\bf r}^2}\right\rangle =\frac 2{\sqrt{\sigma _0} }\left(
\frac{a(n)}3\right) ^{3/4}\left[ 1+\frac{3m_2^2}{8\sigma _0}\left( \frac
3{a(n)}\right) ^{3/2}\right] ^{-1/3}. \label{69}
\end{equation}
For $\rho$-meson putting $m_2=0$ in Eq. (69) we arrive at
\begin{equation}
\left\langle \sqrt{{\bf r}^2}\right\rangle =\frac 2{\sqrt{\sigma _0} }\left(
\frac{a(n)}3\right) ^{3/4}. \label{70}
\end{equation}
The same expression (70) was found in [31] using another method. With the
help of the definition of the center of mass coordinate we can write the
approximate relation for the mean charge radius of $\rho$-mesons \footnote{
The relationship $\sqrt{\left\langle {\bf r}^2\right\rangle }\simeq
\left\langle \sqrt{{\bf r}^2}\right\rangle $ is confirmed by the numerical
calculations}:
\begin{equation}
\sqrt{\left\langle r_{\rho}^2\right\rangle }\simeq \frac{1}{2} \left\langle
\sqrt{{\bf r}^2}\right\rangle. \label{71}
\end{equation}
At $\sigma _0=0.15~GeV^2$ [17,44] and $a(1)=2.3381$ Eqs. (70), (71) give
\begin{equation}
\sqrt{\left\langle r_{\rho}^2\right\rangle }\simeq 0.42~fm~~~\left(
~\left\langle \sqrt{{\bf r}^2}\right\rangle =0.84~fm\right).
\label{72}
\end{equation}
The value (72) characterizes the radius of the sphere where the wave function
of the $\rho$ meson is concentrated (remember that ${\bf r}$ is the distance
between quarks). We know only the experimental data for $\pi^{\pm}$ mesons
which have the same quark structure as $\rho^{\pm}$ mesons [45]:
\[
\left\langle
r_{\pi ^{\pm }}^2\right\rangle_{exp}=(0.44\pm 0.02)~fm^2~~~ \left( \sqrt{
\left\langle r_{\pi ^{\pm }}^2\right\rangle }_{exp}\simeq 0.66~fm\right).
\]
For calculating the relative coordinate of quarks for $K^*$ mesons we should use
Eq. (67) or (69) with the conditions $\mu _1=\mu _u$ and $\mu _2=\mu _s$
(Eqs. (66)). As a result formula (67) gives the value of the mean relative
coordinate of $K^*$ mesons:
\begin{equation}
\left\langle \sqrt{{\bf r}^2}\right\rangle _{K^*}=0.79~fm. \label{73}
\end{equation}
With the help of this relation we can estimate the mean charge radius of $K^*
$ mesons
\begin{equation}
\sqrt{\left\langle r_{K^*}^2\right\rangle }\simeq \frac{\mu _2}{\mu _1+\mu
_2 } \left\langle \sqrt{{\bf r}^2}\right\rangle _{K^*}=0.54\left\langle
\sqrt{{\bf r}^2}\right\rangle _{K^*}=0.43~fm. \label{74}
\end{equation}
The experimental data of the mean charge radius of $K^{\pm }$ mesons having
the analogous quark structure as $K^*$ mesons are
\[
\sqrt{\left\langle
r_{K^{\pm }}^2\right\rangle }=(0.53\pm 0.05)~fm~~~~[46],
\]
\[\left\langle
r_{K^{\pm }}^2\right\rangle =(0.34\pm 0.05)~fm^2~~~~[45],
\] and for neutral $K^0$ mesons [46]
\[\sqrt{\left\langle r_{K^0}^2\right\rangle }=(0.28\pm 0.09)~fm.
\]
Expression (69) gives a value for the charge radius of $K^*$-mesons, that is
close to experimental value for $K^{\pm}$-mesons; experimental data for the
charge radius of $K^{*}$-mesons is not available.

The first perturbative one-gluon exchange contribution to the Hamiltonian
determines the spin-spin correction such as the Breit-Fermi hyperfine interaction
[19]. The spin-spin interaction is important to explain the Nambu-Goldstone
phenomenon which takes place for $l=s=0$ channel.
To take into account spin-spin and spin-orbit interactions one needs to use the general
expression (23) and calculate averaged Wilson integral with insertions [17].

Using the Hamiltonian (44) we estimated in [31, 36] the electromagnetic
polarizabilities of mesons on the basis of the expansion in small electromagnetic
fields ${\bf E}$, ${\bf H}$. The same procedure will be used bellow for nucleons.

\section{Green's Function of Three-Quark System}

Now we consider baryons as a three-quark system.
Let us consider the Lorentz-covariant
and gauge invariant combination of a three quark, colorless system (baryon) [17]
\begin{equation}
X_B(x,y,z,C_i)=\epsilon _{abc}\left[ \Phi (Z_0,x)q(x)\right] _a\left[ \Phi
(Z_0,y)q(y)\right] _b\left[ \Phi (Z_0,z)q(z)\right] _c,  \label{(75)}
\end{equation}
where $q(x)$ is a quark bispinor; $a$, $b$, $c$ are colour indexes so that
\[
\left[ \Phi (Z_0,x)q(x)\right] _a=\Phi _{aa^{^{\prime
}}}(Z_0,x)q_{a^{^{\prime }}}(x)
\]
 and $\epsilon _{abc}$ is the Levi-Civita
symbol ($\epsilon _{123}=1$).
The parallel transporter $\Phi (Z_0,x)$ is given by Eq. (2). The points
$x$, $y$, $z$ are the initial coordinates of three quarks with fields
$q(x)$, $q(y)$ and $q(z)$, and the point $Z_0$ is arbitrary.
 Later, the position of $Z_0$ will be defined by requiring to
have the minimal area for the world surface of three-quark system [17].

The two particle quantum Green function of a baryon is given by [17]:
\begin{equation}
G(xyz,x^{\prime }y^{\prime }z^{\prime })=\langle X_B(x,y,z,C_i)\bar {X}
_B(x^{\prime },y^{\prime },z^{\prime },C_i^{\prime })\rangle,  \label{76}
\end{equation}
were $x^{\prime }$, $y^{\prime }$,$z^{\prime}$ are final positions of three quarks;
$\bar {X}_B(x^{\prime },y^{\prime },z^{\prime },C_i^{\prime })$
corresponds to the final state of a baryon:
\begin{equation}
\bar{X}_B(x^{\prime},y^{\prime},z^{\prime},C_i^{\prime
})=\epsilon _{mnk}\left[ \bar{q}(x^{\prime})\Phi (x^{\prime
},Z_0^{\prime })\right] _m\left[ \bar{q}(y^{\prime })\Phi
(y^{\prime },Z_0^{\prime })\right] _n\left[ \bar{q}(z^{\prime
})\Phi (z^{\prime},Z_0^{\prime })\right] _k,  \label{77}
\end{equation}
where $\bar{q}=q^{+}\gamma _4$.
Using the generating functional (6) the Green function (76) takes the form
\[
G(xyz,x^{\prime }y^{\prime }z^{\prime }) =\int DA_\mu \biggl [\epsilon _{abc}
\epsilon_{mnk}\Phi _{aa^{\prime}}(Z_0,x)\Phi _{bb^{\prime }}(Z_0,y)\Phi
_{cc^{\prime}}(Z_0,z)
\]
\[
\times \Phi _{m^{\prime}m}(x^{\prime },Z_0^{\prime })\Phi
_{n^{\prime}n}(y^{\prime },Z_0^{\prime})\Phi _{k^{\prime
}k}(z^{\prime },Z_0^{\prime})
\]
\begin{equation}
\times \delta ^6/\delta \bar{\eta }_{a^{\prime }}(x)\delta \bar{
\eta }_{b^{\prime }}(y)\delta \bar{\eta }_{c^{\prime }}(z)\delta
\eta _{m^{\prime }}(x^{\prime })\delta \eta _{n^{\prime
}}(y^{\prime })\delta \eta _{k^{\prime }}(z^{\prime })Z[\bar{
\eta },\eta ]\biggr ]_{\eta =\bar{\eta }=0}.  \label{(78)}
\end{equation}
In Eq. (78) $Z_0$ and $Z_0^{^{\prime }}$ are the initial and final
positions of the string junction, respectively. The total surface which
consists of world motions of quarks and the path of the string junction must
be minimal. This requirement defines the path of the string junction [17].
Because the path-integral in Eq. (78) is a Gaussian, it is integrated over quark
fields $\bar{q}$, $q$ .

Taking into account Eq. (9) and calculating the variation derivatives in Eq. (78) we
find the quantum Green function of a baryon:
\[
G(xyz,x^{\prime }y^{\prime }z^{\prime })=\int DA_\mu \det (-\gamma _\mu
D_\mu -m)\exp \left\{ iS(A)\right\} \epsilon _{abc}\epsilon _{mnk}
\]
\[
\times \biggl [S_{bm}^\Phi (y,x^{\prime })S_{an}^\Phi (x,y^{\prime
})S_{ck}^\Phi (z,z^{\prime })-S_{am}^\Phi (x,x^{\prime })S_{bn}^\Phi
(y,y^{\prime })S_{ck}^\Phi (z,z^{\prime })
\]
\begin{equation}
+S_{cm}^\Phi (z,x^{\prime })S_{bn}^\Phi (y,y^{\prime })S_{ak}^\Phi
(x,z^{\prime })-S_{cm}^\Phi (z,x^{\prime })S_{an}^\Phi (x,y^{\prime
})S_{bk}^\Phi (y,z^{\prime })  \label{(79)}
\end{equation}
\[
+S_{am}^\Phi (x,x^{\prime })S_{cn}^\Phi (z,y^{\prime })S_{bk}^\Phi
(y,z^{\prime })-S_{bm}^\Phi (y,x^{\prime })S_{cn}^\Phi (z,y^{\prime
})S_{ak}^\Phi (x,z^{\prime })\biggr ],
\]
where we introduce the following notation for the covariant Green function
\begin{equation}
S_{am}^\Phi (x,x^{\prime })=\Phi _{aa^{\prime }}(Z_0,x)S_{a^{\prime
}m^{^{\prime }}}(x,x^{\prime })\Phi _{m^{\prime }m}(x^{\prime
},Z_0^{\prime }).  \label{(80)}
\end{equation}
As in the case of mesons the functional determinant in Eq. (79) gives
the contribution of the additional quark loops. The presence of
different terms in Eq. (80) is connected with the permutations of quark
fields because the quantum Green function being considered. As different terms
in Eq. (79) have the same structure, we consider in detail only one
term. Neglecting the functional determinant we find the approximate
expression for the baryon Green function
\begin{equation}
G_1(xyz,x^{\prime }y^{\prime }z^{\prime })=-\int DA_\mu \exp \left\{
iS(A)\right\} \epsilon _{abc}\epsilon _{mnk}S_{am}^\Phi (x,x^{\prime
})S_{bn}^\Phi (y,y^{\prime })S_{ck}^\Phi (z,z^{\prime }).
\label{(81)}
\end{equation}
Expression (81) is the basic formula for deriving effective action for baryons.

\section{Effective Action for Baryons}

We consider baryons in external electromagnetic fields.
Inserting Eq. (21) into Eq. (80) we find
\[
S_{am}^\Phi (x,x^{\prime })=-i\int_0^\infty ds\int_{z(0)=x^{\prime
}}^{z(s)=x}Dz\left( m_1-\frac i2\gamma _\mu \dot z_\mu
(t)\right) P_\Sigma
\]
\[
\times \exp \biggl\{ i\int_0^sdt\biggl[ \frac 14\dot z_\mu
^2(t)-m_1^2+e_1\dot z_\mu (t)A_\mu ^{el}(z)
\]
\begin{equation}
+\Sigma _{\mu \nu}(e_1 F_{\mu \nu }^{el}+gF_{\mu \nu})\biggr] \biggr\} \left( \Phi _{C_x}
(Z_0,Z_0^{\prime })\right) _{am},
\label{(82)}
\end{equation}
where the contour $C_x$ in Eq. (82) consists of lines between $Z_0$, $x$
and $Z_0^{\prime }$, $x^{\prime }$ and path $z_\mu (t)$. Using the
expression (82) for each quark, the baryon Green function (81) becomes
\[
G_1(xyz,x^{\prime }y^{\prime }z^{\prime })=i\int_0^\infty \prod_j
ds_j\int \prod_j Dz^{(j)}\left( m_j-\frac i2\gamma _\mu
\dot z_\mu ^{(j)}(t_j)\right)
\]
\[
\times P_\Sigma \prod_j \exp \left\{ \int_0^{s_j}dt_j\Sigma
_{\mu \nu }\frac \delta {\delta \sigma _{\mu \nu }^{(j)}(t_j)}\right\}
\]
\[
\times \exp \biggl\{ i\sum_j \int_0^{s_j}dt_j\biggl[ \frac
14\left( \dot z_\mu ^{(j)}(t_j)\right) ^2-m_j^2+e_j
\dot z_\mu ^{(j)}(t_j)A_\mu ^{el}(z^{(j)})
\]
\begin{equation}
+e_j\Sigma _{\mu \nu }F_{\mu \nu }^{el}\biggr] \biggr\}
 \langle W(C_xC_yC_z)\rangle_A,
 \label{(83)}
\end{equation}
where $j=1,2,3$; $e_j$ is the charge of the $j$-th quark; the boundary
conditions $z_\mu ^{(1)}(0)=x_\mu ^{\prime }$, $z_\mu ^{(1)}(s_1)=x_\mu $
, $z_\mu ^{(2)}(0)=y_\mu ^{\prime }$, $z_\mu ^{(2)}(s_2)=y_\mu $, $z_\mu
^{(3)}(0)=z_\mu ^{\prime }$, $z_\mu ^{(3)}(s_3)=z_\mu $ are used here and
the Wilson loop is given by (see [17])
\[
\langle W(C_xC_yC_z)\rangle_A =\epsilon _{abc}\epsilon _{mnk}\langle \left(
\Phi _{C_x}(Z_0,Z_0^{^{\prime }})\right) _{am}
\]
\begin{equation}
\times\left( \Phi
_{C_y}(Z_0,Z_0^{\prime })\right) _{bn}\left( \Phi
_{C_z}(Z_0,Z_0^{\prime })\right) _{ck}\rangle_A.  \label{(84)}
\end{equation}
We took into account relations such as (28) (see [17]).
Contours $C_x$, $C_y$, $C_z$ correspond to three quarks which have paths $
z_\mu ^{(1)}$, $z_\mu ^{(2)}$, $z_\mu ^{(3)}$ and masses $m_1$, $m_2$, $m_3$,
 respectively. Relationship (83) is the generalization of one [17] for
the case of quarks placed in external electromagnetic fields which
possess spins.

We imply further that the average distance between quarks $\langle r\rangle $ is
greater than the time fluctuations (in units $c=\hbar =1$) of the gluonic
fields $T_g$: $\langle r\rangle >T_g$. This condition is valid not only for
asymptotic baryon states of the Regge trajectories with large angular momenta
of the baryon but also for lower baryon states. The asymptotic of the average
Wilson loop integral obeys then the area law and is given by (in the
Minkowski space):
\begin{equation}
\langle W(C_xC_yC_z)\rangle_A =\exp \left\{ -i\sigma \left( S_1+S_2+S_3\right)
\right\},  \label{(85)}
\end{equation}
where $\sigma $ is the string tension and $S_j$ ($j=1,2,3$) is the minimal
surface bounded by the trajectories of the quarks $q_j$ and string junction
$Z_0$. The path of the string junction is defined by the requirement that the
sum $S_1+S_2+S_3$ is minimal [17]. Following [17], new variables are
introduced:
\begin{equation}
\tau =\frac{t_1T}{s_1}=\frac{t_2T}{s_2}=\frac{t_3T}{s_3},\hspace{0.3in}\mu
_j=\frac T{2s_j},  \label{(86)}
\end{equation}
where $\tau $ means the proper time for every quark and $\mu _j$ ($j=1,2,3$
) is the dynamical mass of the $j-$th quark. Using Eqs. (85), (86) from
Eq. (83) we arrive at
\[
G_1(xyz,x^{\prime }y^{\prime }z^{\prime })=-i\frac{T^3}8\int_0^\infty
\prod_j \frac{d\mu _j}{\mu _j^2}\int \prod_j
Dz^{(j)}\left( m_j-i\mu _j\gamma _\mu \dot z_\mu ^{(j)}(\tau
)\right)
\]
\[
\times P_\Sigma \prod_j \exp \left\{ \frac 1{2\mu _j}\Sigma
_{\mu \nu }\int_0^Td\tau \frac \delta {\delta \sigma _{\mu \nu }^{(j)}(\tau
)}\right\}
\]
\[
\times \exp \biggl\{i\int_0^Td\tau \sum_j \biggl[ \frac 12\mu
_j\left( \dot z_\mu ^{(j)}(\tau )\right) ^2-\frac{m_j^2}{2\mu
_j}+e_j\dot z_\mu ^{(j)}(\tau )A_\mu ^{el}(z^{(j)})
\]
\begin{equation}
+\frac{e_j}{2\mu _j}\Sigma _{\mu \nu }F_{\mu \nu }^{el}\biggr]
-i\sigma \left( S_1+S_2+S_3\right) \biggr\}.  \label{(87)}
\end{equation}
The integral in the last exponential factor of Eq. (87) represents the
effective action for three quark system (baryon) taking into account the
spins of quarks. So $T$ is the time of the observation, $\tau $ is the
proper time of quarks; $m_j$ and $\mu _j$ are the current and dynamical
masses of $j-$th quark. It follows from Eq. (87) that there is
integration over dynamical masses $\mu _j$ to get the Green function of the
three quark system. Pre-exponential factors in Eq. (87) allow us to calculate
in principle the spin-spin and spin-orbital contributions to the effective
action. As a first approximation (see [17]) we neglect the short-range spin
corrections and consider therefore scalar quarks. The terms $e_j\Sigma _{\mu
\nu }F_{\mu \nu }^{el}$ which describe the interaction of the magnetic field
with the spins of quarks will be omitted. With this assumption we arrive at the
effective action for baryons in external electromagnetic fields
\[
B=\int_0^Td\tau \sum_j \biggl[ \frac 12\mu _j\left(
\dot z_\mu ^{(j)}(\tau )\right) ^2-\frac{m_j^2}{2\mu _j}
\]
\begin{equation}
+e_j
\dot z_\mu ^{(j)}(\tau )A_\mu ^{el}(z^{(j)})\biggr] -\sigma \left(
S_1+S_2+S_3\right).
\label{(88)}
\end{equation}
The case when electromagnetic fields are absent was considered in [17]. The
terms which describe the interaction of quarks with electromagnetic fields
are essential for us because we are going to calculate electromagnetic
characteristics of nucleons. It is convenient to introduce new variables
$R_\mu $, $\xi _\mu $ and $\eta _\mu $ instead of $z_\mu ^{(j)}$ in
accordance with relationships [17]:
\[
z_\mu ^{(1)}=R_\mu +\left( \frac{\mu \mu _3}{M\left( \mu _1+\mu _2\right)}
\right) ^{1/2}\xi _\mu -\left( \frac{\mu \mu _2}{\mu _1\left( \mu _1+\mu
_2\right) }\right) ^{1/2}\eta _\mu,
\]
\[
z_\mu ^{(2)}=R_\mu +\left( \frac{\mu \mu _3}{M\left( \mu _1+\mu _2\right)}
\right) ^{1/2}\xi _\mu +\left( \frac{\mu \mu _1}{\mu _2\left( \mu _1+\mu
_2\right) }\right) ^{1/2}\eta _\mu,
\]
\begin{equation}
z_\mu ^{(3)}=R_\mu -\left( \frac{\mu \left( \mu _1+\mu _2\right) }{M\mu _3}
\right) ^{1/2}\xi _\mu,  \label{(89)}
\end{equation}
where $R_\mu $ is the center of mass coordinate of a baryon; $\xi _\mu $ and
$\eta _\mu $ are relative coordinates of quarks, $M=\mu _1+\mu _2+\mu _3$ is
the sum of dynamical masses of quarks. The arbitrary mass parameter $\mu$
in Eq. (89) defines the scale of relative coordinates $\xi _\mu $ and
$\eta _\mu $. After the substitution (89), the measure $\prod_j
Dz^{(j)}$ transforms into $DRD\eta D\xi $ in path-integral (87).

Let us consider the uniform and constant external electromagnetic fields.
Then the vector-potential of electromagnetic fields can be represented as
\[
A_\nu^{el}(z^{(j)})=\frac 12 F_{\mu \nu }^{el}z_\mu ^{(j)}.
\]
Inserting Eqs. (89) into Eq. (88) we find the effective action
for the three quark system (see [37]) in the form
\begin{equation}
B=\int_0^Td\tau \left[ \frac M2\dot R_\nu^2+\frac \mu 2\left(
\dot \xi _\nu ^2+\dot \eta _\nu ^2\right) -
\sum_j \frac{m_j^2}{2\mu _j}\right] -\sigma \left(
S_1+S_2+S_3\right) +\Delta B,  \label{(90)}
\end{equation}
\[
\Delta B=\frac 12F_{\nu \mu }^{el}\int_0^Td\tau \biggl [e\dot R
_\mu R_\nu + \lambda \dot \xi _\mu \xi _\nu +\rho
\dot \eta _\mu \eta _\nu+\gamma \left( \dot R_\mu \xi _\nu
+\dot \xi _\mu R_\nu \right)
\]
\begin{equation}
+\delta \left( \dot R_\mu \eta _\nu +
\dot \eta _\mu R_\nu \right) +\delta \sqrt{\frac{\mu \mu _3}{
\left( \mu _1+\mu _2\right) M}}\left( \dot \xi _\mu \eta _\nu
+ \dot \eta _\mu \xi _\nu \right) \biggr ],
 \label{(91)}
\end{equation}
where we introduce parameters:
\[
\gamma =\sqrt{\frac \mu M}\left[ \left( e_1+e_2\right) \sqrt{\frac{\mu _3}{
\mu _1+\mu _2}}-e_3\sqrt{\frac{\mu _1+\mu _2}{\mu _3}}\right],
\]
\[
\delta =\sqrt{\frac \mu {\mu _1+\mu _2}}\left( e_2\sqrt{\frac{\mu _1}{\mu _2}
}-e_1\sqrt{\frac{\mu _2}{\mu _1}}\right),
\]
\[
\rho =\frac \mu {\mu _1+\mu _2}\left( \frac{e_1\mu _2}{\mu _1}+\frac{e_2\mu
_1}{\mu _2}\right),
\]
\begin{equation}
\lambda =\frac \mu M\left[ \frac{\left( e_1+e_2\right) \mu _3}{\mu _1+\mu _2}
+\frac{e_3\left( \mu _1+\mu _2\right) }{\mu _3}\right],  \label{(92)}
\end{equation}
and $e=e_1+e_2+e_3$ is the charge of a baryon. As a particular case, when
electromagnetic fields are absent ($\Delta B=0$) we arrive at the action
derived in [17]. It follows from Eq. (90) that the center of mass
coordinate $R_\mu $ is separated from relative coordinates $\xi _\mu $ and
$\eta _\mu $ and $\mu $ plays the role of the mass of the $\xi _\mu $, $\eta
_\mu $ excitations. Following [17], the straight line approximation for
strings and the asymmetric quark-diquark structure of baryons will be
assumed. The asymmetric configuration (see also [35,38]) means that two
quarks $q^{(1)}$ and $q^{(2)}$ are near each other and quark $q^{(3)}$ is
farther from them. This case is preferable [35,38] and the slope of
linear baryon Regge trajectories is the same as for mesons [17]. Then
$\sqrt{{\bf \xi}^2}\gg \sqrt{{\bf \eta}^2}$ (${\bf \xi}^2 =
\xi _1^2+\xi _2^2+\xi _3^2$) and the coordinate $\eta _\mu $ can be
ignored. We neglect therefore the surfaces $S_1$, $S_2$ and assume for $S_3$
the following expression [17]:
\begin{equation}
S_3=b\int_0^Td\tau \sqrt{{\bf \xi}^2},  \label{(93)}
\end{equation}
where
\begin{equation}
b=\sqrt{\frac{\mu \left( \mu _1+\mu _2\right) }{M\mu _3}}+\sqrt{\frac{\mu
\mu _3}{M\left( \mu _1+\mu _2\right) }}.  \label{(94)}
\end{equation}
Equation (93) takes into account the confinement of quarks and gives the
linear potential between quarks. Using the definition $B=\int_0^Td\tau
\mathcal{L}$, where $\mathcal{L}$ is the Lagrangian, from Eq. (90) we
arrive at the effective Lagrangian for baryons
\begin{equation}
{\cal L}_{eff}=\frac M2\dot R_{\nu}^2+\frac \mu 2
\dot \xi _\nu ^2-\sum_j \frac{m_j^2}{2\mu _j}-\sigma b\sqrt{{\bf \xi}^2}
 +{\cal L}^{el}.  \label{(95)}
\end{equation}
Here we neglect the coordinate $\eta _\mu $ and introduce the notation:
\begin{equation}
{\cal L}^{el}=\frac 12F_{\nu \mu }^{el}\left[ e\dot R_\mu
R_\nu +\lambda \dot \xi _\mu \xi _\nu +\gamma \left(
\dot R_\mu \xi _\nu +\dot \xi _\mu R_\nu \right) \right],
\label{(96)}
\end{equation}
Lagrangian ${\cal L}^{el}$ describes the electromagnetic interaction of
the string. It follows from Eq. (89) that
\begin{equation}
z_\mu ^{(3)}-\frac 12\left( z_\mu ^{(1)}+z_\mu ^{(2)}\right) =-b\xi _\mu
+\frac 12\sqrt{\frac \mu {\mu _1+\mu _2}}\left( \frac{\mu _2-\mu _1}{\sqrt{
\mu _1\mu _2}}\right) \eta _\mu.  \label{(97)}
\end{equation}
As the second term in Eq. (97) is small, the coordinate $\xi _\mu $ is
proportional to the ``distance'' between quark $q^{(3)}$ and the center of
mass of quarks $q^{(1)}$ and $q^{(2)}$ which form a diquark. At the large
time $T$ limit $\xi _4=0$, $R_4=i\tau $ [17] and Lagrangian (96) takes
the form
\begin{equation}
{\cal L}^{el}=e\left( {\bf RE}\right) +\gamma \left( {\bf \xi E}
\right) -\frac 12\epsilon _{mnk}H_k\left( e\dot R_m R_n+\lambda
\dot \xi _m \xi _n+\gamma \dot (R_m \xi _n+
\dot \xi _m R_n)\right),  \label{(98)}
\end{equation}
where the electric field $E_k=iF_{k4}$ and magnetic field $H_k=(1/2)\epsilon
_{kmn}F_{mn}$. To clarify the physical meaning of the terms in Eq. (98),
let us consider the dipole moment of quarks. Using the definition of the
electric dipole moment and Eq. (89) we have
\begin{equation}
{\bf d}=\sum_j e_j{\bf z}^{(j)}=e{\bf R}+\gamma
{\bf \xi }+\delta {\bf \eta }.  \label{(99)}
\end{equation}
So, first two terms in Eq. (98) ( neglecting the coordinate $\eta _\mu $
) describe the interaction of the dipole moment of quarks with the electric
field in accordance with the expression for the potential energy: $U=-(
{\bf dE})$. The magnetic moment is given by
\[
m_k=\frac 12\epsilon _{mnk}\sum_j e_jz_m^{(j)}\dot
z_n^{(j)}=\frac 12\epsilon _{mnk}\biggl [eR_m\dot R
_n+\lambda \xi _m\dot \xi _n+\gamma \left( R_m\dot
\xi _n+\xi _m\dot R_n\right)
\]
\begin{equation}
+\rho \eta _m\dot \eta _n+\delta \left( R_m\dot
\eta _n+\eta _m\dot R_n\right) +\delta \sqrt{\frac{\mu \mu _3
}{\left( \mu _1+\mu _2\right) M}}\left( \xi _m\dot \eta
_n+\eta _m\dot \xi _n\right) \biggr ].  \label{(100)}
\end{equation}
It follows from Eqs. (91), (98) that there is an interaction of the
magnetic field with the magnetic moment in such a way that the interaction
energy is $U=-\left( {\bf mH}\right)$. So Lagrangian (98) describes
the interaction of electric and magnetic moments of baryons with electric
and magnetic fields, respectively.

\section{Mean Relative Coordinates of quarks in Nucleons}

Now we estimate the mean size of baryons.
At the large time $T$ limit $\stackrel{\cdot }{\xi }_0=0$ ($\xi _4=i\xi _0)$,
 $R_4=i\tau $ [17] and therefore only three dimensional quantities $R_k$
and $\xi _k$ are important. From Eq. (98) we find three momenta
corresponding to the center of mass coordinate $R_k$ and relative coordinate
$\xi _k:$
\[
\Pi _k=\frac{\partial {\cal L}_{eff}}{\partial \dot R_k}=M
\dot R_k-\frac 12\epsilon _{knm}H_m\left( eR_n+2\gamma \xi
_n\right),
\]
\begin{equation}
\pi _k=\frac{\partial {\cal L}_{eff}}{\partial \dot \xi _k}
=\mu \dot \xi _k-\frac 12\epsilon _{knm}H_m\left( \lambda \xi
_n+2\gamma R_n\right).  \label{(101)}
\end{equation}
Here we take into account that Lagrangian is defined within the accuracy of
the total derivative on time. Then the effective Hamiltonian ${\cal H}
_{eff}=\Pi _k\dot R_k+\pi _k\dot \xi _k-{\cal
 L}_{eff}$ corresponding to the quark-diquark structure of a baryon is
given by
\begin{equation}
{\cal H}_{eff}=\sum_j \frac{m_j^2}{2\mu _j}+\frac M2+\frac
M2\dot R_k^2+\frac \mu 2\dot \xi _k^2+\sigma
b \sqrt {{\bf \xi}^2} -e\left( {\bf ER}\right) -\gamma \left( {\bf
E\xi }\right),  \label{(102)}
\end{equation}
The Hamiltonian for baryons Eq. (102) looks like the one for mesons (44) because we
consider basically the string between quark and diquark. Therefore the
calculations of mean coordinates and electromagnetic polarizabilities is
the same [31,36]. But in the case of baryons there are more parameters and the
analysis is more complicated.

With the help of Eq. (101) the effective Hamiltonian (102) is rewritten as
\[
{\cal H}_{eff}=\sum_j \frac{m_j^2}{2\mu _j}+\frac M2+\frac
1{2M}\left[ {\bf \Pi} +\frac e2\left( {\bf R}\times {\bf H}\right)
+\gamma \left( {\bf \xi }\times {\bf H}\right) \right] ^2
\]
\begin{equation}
+\frac 1{2\mu }\left[ {\bf \pi }+\frac \lambda 2\left( {\bf \xi }
\times {\bf H}\right) +\gamma \left( {\bf R}\times {\bf H}\right)
\right] ^2+\sigma b \sqrt {{\bf \xi}^2} -e\left( {\bf ER}\right)
-\gamma \left( {\bf E\xi }\right),  \label{(103)}
\end{equation}
where $\left( {\bf \xi }\times {\bf H}\right) _k=\epsilon _{mnk}\xi
_mH_n$. The mass of a baryon ${\cal M}(\mu _j)$ is defined here as a
solution to equation
\begin{equation}
{\cal H}_{eff}\Phi ={\cal M}(\mu _j)\Phi.  \label{(104)}
\end{equation}
In according to the Noether theorem, the momentum ${\bf \Pi }$ is
conserved and we can put ${\bf R}={\bf \Pi }=0$ in Eq. (103). To find the
solution to Eq. (104) we can use the substitution $\pi _k=-i\partial
/\partial \xi _k$. Then the mass of a baryon ${\cal M}(\mu _j)$ is given
by
\begin{equation}
{\cal M}(\mu _j)=\sum_j \frac{m_j^2}{2\mu _j}+\frac
M2+\epsilon (\mu _j,{\bf E,H}),  \label{(105)}
\end{equation}
where $\epsilon (\mu _j,{\bf E,H})$ is the eigenvalue of the equation
\[
\left\{ \frac 1{2\mu }\left[ -i\frac {\partial} {\partial {\bf \xi
}}-\frac \lambda 2\left( {\bf \xi }\times {\bf H}\right) \right] ^2+
\frac{\gamma ^2}{2M}\left( {\bf \xi }\times {\bf H}\right) ^2+\sigma
b\sqrt{{\bf \xi}^2} -\gamma \left( {\bf E\xi }\right) \right\} \Phi
\]
\begin{equation}
=\epsilon (\mu _j,{\bf E,H})\Phi.  \label{(106)}
\end{equation}
The term $(\gamma ^2/(2M))\left( {\bf \xi }\times {\bf H}\right) ^2$
in Eq. (106) is due to the recoil of the string. In nonrelativistic models
the effect of the recoil was studied in [39] (see also [22]). As we
neglected the spin of baryons here, there is no interaction of spin with the
external magnetic field. It is not difficult to take into account such
interaction. Eq. (106) is like the equation for mesons [36].

If ${\bf E}={\bf H}=0$, we arrive from Eq. (106) at
\begin{equation}
\left( -\frac 1{2 \mu }\frac{\partial ^2}{\partial \xi_i^2}+\sigma
b \sqrt{{\bf \xi}^2} \right) \Phi =\epsilon (\mu_j )\Phi. \label{107}
\end{equation}
Equation (107) has the same structure as Eq. (51).
The eigenvalue of Eq. (107) is
\begin{equation}
\epsilon (\mu _j)=\left( 2\mu \right) ^{-1/3}\left( b\sigma \right)
^{2/3}a(n)=(\sigma )^{2/3}a(n)\left[ \frac M{2\mu _3\left( \mu _1+\mu
_2\right) }\right] ^{1/3}.  \label{(108)}
\end{equation}
The condition of the minimum of the baryon mass (105) ($\partial {\cal M}
(\mu _j)/\mu _j=0$) at $m_1=m_2$ (and $\mu _1=\mu _2$), with the help of
Eq. (105) gives the dynamical mass of a diquark:
\begin{equation}
\mu _3^{(0)}=\mu _1^{(0)}+\mu _2^{(0)}=\sqrt{\sigma }\left[ \frac{a(n)}
3\right] ^{3/4}.  \label{(109)}
\end{equation}
This value is different from one [17] obtained for large angular momentum.
Using Eq. (109) we arrive from Eq. (105) at the expression for the mass of
a baryon (see also [17]):
\begin{equation}
{\cal M}(\mu _j)=\frac{m_3^2+4m_1^2}{2\mu _3^{(0)}}+4\mu _3^{(0)}.
\label{(110)}
\end{equation}
To estimate the baryon mass, the value of the string tension $\sigma =0.15$
GeV$^2$ will be used [17]. Neglecting the small current masses of quarks
$m_j$ we find from Eqs. (109), (110) the mass of a diquark for $n=1$: $\mu
_3^{(0)}=320$ MeV and the nucleon mass : ${\cal M}(\mu _j)=1.28$ GeV [37].
This value of the nucleon mass is a little greater then real nucleon mass
because spin-spin and spin-orbit forces were omitted.

Now we consider the mean relative coordinates of nucleons on the basis of the
virial theorem which gives (as for the case of mesons [36]) the mean potential
energy $\langle U\rangle =2\langle T\rangle $, where $\langle T\rangle $ is
the mean kinetic energy. Then using the relation $\langle T\rangle +\langle
U\rangle =\epsilon (\mu _j),$ we arrive at
\begin{equation}
\langle U\rangle =\frac 23\epsilon (\mu _j)=\frac 23(2\mu )^{-1/3}(b\sigma
)^{2/3}a(n).  \label{(111)}
\end{equation}
Comparing Eq. (111) with the relation $\langle U\rangle =b\sigma \langle
\sqrt{{\bf \xi }^2}\rangle $ gives the following expression
\begin{equation}
\langle \sqrt{{\bf \xi }^2}\rangle =\frac 23(2\mu b\sigma
)^{-1/3}a(n)=\sigma ^{-1/4}\sqrt{\frac 2\mu }\left( \frac{a(n)}3\right)
^{9/8}.  \label{(112)}
\end{equation}
In accordance with Eq. (97) the size of the nucleon is characterized by
the value $\mid {\bf z}^{(3)}-(1/2)({\bf z}^{(1)}+{\bf z}
^{(2)})\mid \simeq \mid b{\bf \xi }\mid $. Introducing the notation $
{\bf r}_b=b{\bf \xi }$, from Eq. (112) we have
\begin{equation}
\langle \sqrt{{\bf r}_b^2}\rangle =\frac 2{\sqrt{\sigma }}\left( \frac{a(n)}
3\right) ^{3/4}.  \label{(113)}
\end{equation}
The same expression was found for mesons (see Eq. (70)). For the quark-diquark
system the string tension coincides with those of mesons and therefore the
quark-diquark system has approximately the same size as mesons. Using $
\sigma =0.15$ GeV$^2$ and $a(1)=2.2281$ we find the mean size of the
nucleons
\begin{equation}
\langle \sqrt{{\bf r}_b^2}\rangle =0.84~fm.  \label{(114)}
\end{equation}
The experimental value of the charge radii of the proton and neutron are
\[
\sqrt{\langle r_p^2\rangle }=0.86~ fm ~~~~[47],
\]
\[
\sqrt{\langle r_n^2\rangle }
=(-0.113\pm 0.003)~ fm ~~~~[48].
\]
In accordance with Eq. (89) the center of mass
of the quark-diquark system is situated in the center between quark $q^{(3)}$
and diquark $\left( q^{(1)},q^{(2)}\right) $ and the mean radius of a
nucleon is $\left( 1/2\right) \langle \sqrt{{\bf r}_b^2}\rangle =0.42$ fm
which is the reasonable value. It should be noted that charge radii of
hadrons are defined from electromagnetic formfactors.

\section{ Electric Polarizabilities of Nucleons}

To evaluate electric polarizabilities of nucleons we consider the case when
${\bf H}=0$, ${\bf E}\neq 0$. It is
possible to assume as an approximation that ${\bf E}\parallel {\bf \xi }$ [37],
i.e. external electric field is parallel to the string which connects quark
$q^{(3)}$ with diquark $\left( q^{(1)},q^{(2)}\right) $. So we neglect the
rotation of the string. It is justified only for the ground state when the
orbital quantum number $l=0$. Introducing the effective string tension
\begin{equation}
\sigma _{eff}=\sigma -\frac \gamma bE,  \label{(115)}
\end{equation}
we arrive from Eq. (108) at the eigenvalue
\[
\epsilon (\mu _j,{\bf E}
)=\left( 2\mu \right) ^{-1/3}\left( b\sigma _{eff}\right) ^{2/3}a(n).
\]
 From (109), (110), by neglecting the small terms containing the current masses
we find the mass of a baryon
\begin{equation}
{\cal M}(\mu _j,{\bf E})=4\sqrt{\sigma _{eff}}\left[ \frac{a(n)}
3\right] ^{3/4},  \label{(116)}
\end{equation}
which depends on the external electric field. Inserting Eq. (115) into
Eq. (116) and expanding it in a small electric field one yields
\begin{equation}
{\cal M}(\mu _j,{\bf E})\simeq \left[ \frac{a(n)}3\right] ^{3/4}\left(
4\sqrt{\sigma }-\frac q{\sqrt{\sigma }}E-\frac{q^2}8\sigma ^{-3/2}E^2\right),
  \label{(117)}
\end{equation}
where Eq. (109) was used and $q=e_1+e_2-e_3$. We write here only terms of
the expansion ${\cal M}(\mu _j,{\bf E})$ in small electric field up to
$E^2$. The first term in Eq. (117) gives the mass of a baryon. The second one is
connected with the potential energy of a dipole moment of quarks in the
external electric field $U=-{\bf dE}$. From Eq. (99) when the center of
mass coordinate ${\bf R}=0$ and $\mid {\bf \xi }\mid $ $\gg \mid
{\bf \eta }\mid $, the dipole moment of quark-diquark system is ${\bf
d }\simeq \gamma {\bf \xi }=(\gamma /b){\bf r}_b$. Comparing the
potential energy of a dipole $U=-(\gamma /b)r_bE$ (at ${\bf E\parallel r}$)
with the second term of Eq. (117): $-(q/\sqrt{\sigma })[a(n)/3]^{3/4}E$ we
arrive at the expression for the mean relative coordinate $r_b=(2/\sqrt{\sigma
} )[a(n)/3]^{3/4}$ which coincides with Eq. (113). Here we define the
electric dipole moment of quark-diquark system more precisely as compared
with the letter [37] and as a result the mean relative coordinate of a baryon
Eq. (113) coincides with that for a meson. The third term in Eq. (117)
describes the potential energy due to the electric polarizability of a
baryon (see Eq. (155) in Appendix).

 From Eq. (117) by comparing the quadratic term in
${\bf E}$ with Eq. (155) we arrive at the electric polarizability of a
baryon:
\begin{equation}
\alpha =\frac{q^2}4\left[ \frac{a(n)}3\right] ^{3/4}\sigma ^{-3/2}.
\label{(118)}
\end{equation}
Let us consider the estimation of the electric polarizability for the proton
$p=uud$. There are two possibilities for a proton as a quark-diquark system:
a) the quark $q^{(3)}=d$ and diquark $(q^{(1)}q^{(2)})=(uu)$, so the
electric charges $e_1=e_2=(2e)/3$, $e_3=-e/3$ and parameter $
q=e_1+e_2-e_3=(5e)/3$; b) the quark $q^{(3)}=u$, diquark $
(q^{(1)}q^{(2)})=(ud)$ and the electric charges $e_1=e_3=(2e)/3$, $e_2=-e/3$
and parameter $q=e_1+e_2-e_3=-e/3$. It should be noted that there are no
permutations of quarks here which occur in the Green function Eq. (79) due
to the Pauli principle. Using the value of the string tension $\sigma =0.15$
GeV$^2$ and $a(1)=2.3381$ from Eq. (118) we find the static polarizability
of a proton in Gaussian units for two cases
\begin{equation}
a)~ \alpha _p=5.56\times 10^{-4}~ fm^3,\hspace{0.3in}
b)~ \alpha _p=0.22\times 10^{-4}~ fm^3.  \label{(119)}
\end{equation}
From Eq. (161) (see Appendix) $\Delta \alpha_p=(4.5\pm 0.1)\times 10^{-4}~ fm^3$,
and the total electric polarizability of a proton becomes (see Eq. (158))
\begin{equation}
a)~ \bar{\alpha }_p=10\times 10^{-4}~  fm^3,
\hspace{0.3in}b)~ \bar{\alpha }_p=4.7\times 10^{-4}~ fm^3.
\label{(120)}
\end{equation}
The configuration of a proton a), when the quark $q^{(3)}=d$ and diquark
$(q^{(1)}q^{(2)})=(uu)$ is more favorable as the experimental values of
electric polarizability are
\[
\bar{\alpha }_p^{\exp } =\left( 10.9\pm
2.2\pm 1.3\right) \times 10^{-4}~ fm^3~~~~ [49],
\]
\[
\bar{\alpha }_p^{\exp }=\left( 10.6\pm 1.2\pm
1.0\right) \times 10^{-4}~ fm^3~~~~ [50],
\]
\[
\bar{\alpha }_p^{\exp }=(9.8\pm 0.4\pm 1.1)\times 10^{-4}~ fm^3~~~~ [51].
\]
 So in the case a) we have a good agreement with experimental
data.

For the neutron $n=udd$  there are also two possibilities: a) the quark
$q^{(3)}=u$ and diquark $(q^{(1)}q^{(2)})=(dd)$, so the electric charges
$e_1=e_2=-e/3$, $e_3=(2e)/3$ and parameter $q=e_1+e_2-e_3=-(4e)/3$; b) the
quark $q^{(3)}=d$, diquark $(q^{(1)}q^{(2)})=(ud)$ and the electric charges
$e_3=e_2=-e/3$, $e_1=(2e)/3$ and parameter $q=e_1+e_2-e_3=(2e)/3$. Inserting
these parameters into Eq. (118) one gives
\begin{equation}
a)~ \alpha _n=3.56\times 10^{-4}~ fm^3,\hspace{0.3in}
b)~ \alpha _n=0.89\times 10^{-4}~ fm^3.  \label{(121)}
\end{equation}
For the neutron, in accordance with Eq. (161) $\Delta \alpha _n=0.62\times 10^{-4}$ fm$^3$
and the generalized electric polarizability of a neutron is given by
\begin{equation}
a)~ \bar{\alpha }_n=4.2\times 10^{-4}~ fm^3,
\hspace{0.3in}b)~ \bar{\alpha }_n=1.5\times 10^{-4}~ fm^3.
\label{(122)}
\end{equation}
The experimental situation for a neutron is more complicated as there are
different experimental data:
\begin{equation}
\bar{\alpha }_n^{\exp } =(0.0\pm 5)\times
10^{-4}~ fm^3~~~~ [48],
\label{(123)}
\end{equation}
\begin{equation}
\bar{\alpha }_n^{\exp }=(12.6\pm 1.5\pm 2.0)\times 10^{-4}~ fm^3~~~~ [52].
\label{(124)}
\end{equation}
The recent experimental data (123) are close to both cases but the case b) with
the quark $q^{(3)}=d$ and diquark $(q^{(1)}q^{(2)})=(ud)$ is more favorable.
Other experimental data [52] are close to the case a) in Eq. (122) with
quark $q^{(3)}=u$ and diquark $(q^{(1)}q^{(2)})=(dd)$ but the magnitude
(122) a) is small. This experimental discrepancy does not allow us to choose
reliably one of the possibilities: a) or b). The value of $\bar{\alpha}
_n$ in the situation a) in Eq. (122) is close to one obtained in the
oscillator nonrelativistic quark model [53-55,22]. The results of calculations in the
framework of the dispersion sum rule and CHPT Eq. (167) are closer to the case a).

\section{Diamagnetic Polarizabilities of Nucleons}

In according to Eq. (103) for calculating the magnetic polarizability of
nucleons one needs to compare it with Eq. (155). The effective Hamiltonian
(103) (at ${\bf E}=0$, ${\bf R}={\bf \Pi }=0$) can be cast into
\[
{\cal H}_{eff}={\cal H}_0+{\cal H}_{int},
\]
\[
{\cal H}_0=\sum_j \frac{m_j^2}{2\mu _j}+\frac M2-\frac
1{2\mu }\frac{\partial ^2}{\partial \xi _j^2}+\sigma b \sqrt{{\bf \xi}^2},
\]
\begin{equation}
{\cal H}_{int}=-\frac \lambda {2\mu }{\bf HL}+\left( \frac{\lambda ^2}{
8\mu }+\frac{\gamma ^2}{2M}\right) \left[ \left( {\bf \xi }\times {\bf
 H}\right) \right] ^2,  \label{(125)}
\end{equation}
where $L_k=-i\epsilon _{kmn}\xi _m\partial _n$ is the angular momentum and
$\partial _n$ =$\partial /\partial \xi _n$. Without loss of generality we can
choose the direction of the magnetic field on the third axis, i.e. ${\bf
H }=(0,0,H).$ Then the Hamiltonian of an interaction of quarks with the
magnetic field is given by
\begin{equation}
{\cal H}_{int}=-\frac{\lambda H}{2\mu }L_3+H^2\left( \frac{\lambda ^2}{
8\mu }+\frac{\gamma ^2}{2M}\right) \left( \xi _1^2+\xi _2^2\right),
\label{(126)}
\end{equation}
where $L_3=i\left( \xi _2\partial _1-\xi _1\partial _2\right) $. Considering
the small external magnetic field, the perturbative theory can be applied.
Using the perturbative method [56], within the accuracy of the second order,
one arrives at the shift of the energy
\[
\Delta {\cal E}_n=\langle n\mid \left[ -\frac{\lambda H}{2\mu }L_3+\left(
\frac{\lambda ^2}{8\mu }+\frac{\gamma ^2}{2M}\right) H^2\xi ^2\sin \vartheta
\right] \mid n\rangle
\]
\begin{equation}
+\sum_{n^{\prime}}  \frac{\mid\langle n^{\prime }\mid -\left(
\lambda HL_3\right) /\left( 2\mu \right) \mid n\rangle\mid^2 }{{\cal E}_n-
{\cal E}_{n^{\prime }}},  \label{(127)}
\end{equation}
where $\vartheta $ is the angle between coordinate ${\bf \xi }$ and
magnetic field ${\bf H}$. If the first term in Eq. (127) is not equal to
zero then the second and third terms are smaller, and the main contribution
to the energy comes from the first term. This occurs when the orbital quantum number
$l>0$. In the case $l=0$, the shift of the energy due to the interaction with
the magnetic field is defined by the second term in Eq. (127).

After averaging in Eq. (127) and taking into
account the equation
\[
\frac {1}{4\pi} \int \sin ^2\vartheta d\Omega = \frac 23
\]
 we find for the ground state when $l=0$, the
shift of energy
\begin{equation}
\Delta {\cal E}_2=\left( \frac{\lambda ^2}{4\mu }+\frac{\gamma ^2}
M\right) \frac{H^2}3 \langle {\bf \xi }^2\rangle,  \label{(128)}
\end{equation}
where $\langle {\bf \xi }^2\rangle=\langle 0\mid {\bf \xi }^2 \mid 0\rangle$,
and $\mid 0\rangle$ means the wave function of the ground $s$-state. We took
into account that the first and the third terms in Eq. (127) equal zero because
$L_3\mid 0\rangle =0$.
Here we ignore the spin interactions of a baryon with the external magnetic field
and therefore only diamagnetic polarizability can be defined from Eq. (128).
Using the definition of the relative coordinate ${\bf r}_b=b{\bf
 \xi }$ and comparing Eq. (128) with Eq. (155) gives the diamagnetic
polarizability of a baryon
\begin{equation}
\beta ^{dia}=-\left( \frac{\lambda ^2}{4\mu }+\frac{\gamma ^2}M\right) \frac
2{3b^2}\langle {\bf r}_b^2\rangle.  \label{(129)}
\end{equation}
As required, the diamagnetic polarizability is negative and Eq. (129)
is like the Langevin formula for the magnetic susceptibility of atoms. The
similar expression was derived in [36] for mesons. Taking into account Eqs. (92),
(109), expression (129) is rewritten as
\begin{equation}
\beta ^{dia}=-\left( \frac{e^2}4+q^2\right) \frac{\langle {\bf r}_b
^2\rangle }{6M},  \label{(130)}
\end{equation}
where $M=\mu _1^{(0)}+\mu _2^{(0)}+\mu _3^{(0)}=2\mu _3^{(0)}=2\sqrt{\sigma}
\left[ a(n)/3\right] ^{3/4}$. To calculate the value of $\beta ^{dia}$ for
nucleons we can use the theoretical magnitude of the mean-squared relative
coordinate $\langle {\bf r}_b^2\rangle $ or experimental data for the size
of a nucleon. The first way is preferable. Taking into account Eq. (113),
the value of $M$ and using the approximate relation $\langle {\bf
\ r}_b^2\rangle \simeq \langle \sqrt{{\bf r}_b^2}\rangle ^2,$ Eq. (130)
transforms into
\begin{equation}
\beta ^{dia}=-\frac{e^2+4q^2}{12\sigma ^{3/2}}\left[ \frac{a(n)}3\right]
^{3/4}.  \label{(131)}
\end{equation}
For a proton with the more favorable configuration a), when the quark $
q^{(3)}=d$ and diquark $(q^{(1)}q^{(2)})=(uu)$, parameter $q=(5e)/3$ and Eq. (131)
(at $\sigma =0.15$ GeV$^2$) takes the value
\begin{equation}
\beta _p^{dia}=-8\times 10^{-4}~ fm^3.
\label{(132)}
\end{equation}
This quantity is greater than that found in the nonrelativistic quark model
[22].
To calculate $\beta ^{para}$ in the present approach one needs to take into account the
interaction of spins of quarks with the magnetic field.
For estimation of
the total magnetic polarizability of a proton we use the contribution $\beta _\Delta
^{para}$ and in accordance with Eqs. (159), (162) we find
\begin{equation}
\bar{\beta }_p=\left( 5\pm 3\right) \times 10^{-4}~ fm^3.
\label{(133)}
\end{equation}
 This quantity is close (within two standard deviations) to the experimental value
[52] which was extracted from the measurements of the cross sections of the Compton
scattering on hydrogen:
\begin{equation}
\bar{\beta}_p^{\exp}=\left( 2.9\mp 0.7\mp 0.8\right)\times 10^{-4}~ fm^3.
\label{(134)}
\end{equation}
The value (133) is also in agreement with quantity (167) found in other schemes.
The value $\bar{\beta }_p$ is small due to the partial cancellation of the
positive paramagnetic polarizability $\bar{\beta }_p^{para}$ with the negative
diamagnetic polarizability $\bar{\beta }_p^{dia}$. The theoretical evaluation,
interpretation and experimental extraction of the total magnetic polarizability
$\bar{\beta }_p$ is difficult because it is small. That is why this quantity
is model dependent in different schemes.  The approach considered allows us to
improve the accuracy by using the perturbation in the spin interaction. The next
step is to take into account such corrections.

For a neutron with the configuration a) where the quark $q^{(3)}=u$ and
diquark $(q^{(1)}q^{(2)})=(dd)$, the parameter $q=-(4e)/3$ and Eq. (131)
gives
\begin{equation}
\beta _n^{dia}=-5.4\times 10^{-4}~ fm^3.  \label{(135)}
\end{equation}
Using the paramagnetic polarizability $\beta _\Delta ^{para}=\left( 13\pm
3\right) \times 10^{-4}$ fm$^3$ [57] for a neutron (see Eq. (162)) we get
from Eq. (135) the generalized magnetic polarizability of a neutron
\begin{equation}
\bar{\beta }_n=\left(7.6\pm 3\right) \times 10^{-4}~ fm^3
 \label{(136)}
\end{equation}
which is in agreement within two standard deviations with the
experimental quantity [52]
\[
\bar {\beta}_n^{\exp}=\left( 3.2\mp 1.5\mp 2.0\right)
\times 10^{-4}~ fm^3.
\]
According to Eqs. (120), (122), (133), (136) the sum of nucleon polarizabilities
in our approach are given by
\[
\bar{\alpha}_p+\bar{\beta }_p=\left(15\pm 3\right) \times 10^{-4}~ fm^3,
\]
\begin{equation}
\bar{\alpha}_n+\bar{\beta }_n=\left(11.8\pm 3\right) \times 10^{-4}~ fm^3.
 \label{(137)}
\end{equation}
Values (137) are close to the reliable results found on the basis sum rule (166).
As pion degrees of freedom play a very important role in the nucleon electric
polarizability, one needs to describe pions in this scheme.
In the approach considered here the accuracy of the values of electromagnetic
polarizabilities can be improved by taking into consideration spin
interactions as perturbations in accordance with Eq. (87).

\section{Cluster Expansion}

In this paragraph we go into Euclidean formalism and follow closely the works
[16,17,61,66].
Some important characteristics of strong interactions depend on the Wilson integral
averaged with certain weights. So, spin-spin and spin-orbit quark interactions
can be expressed through the Wilson loop. It is difficult to calculate the Wilson
loop integral in the general case of arbitrary contours due to the nonlinear character
of gluonic fields. However, it is possible to use the powerful method of cluster
expansion [64]. To use this method it is necessary to accept the stochastic nature
of the QCD vacuum [16]. Lattice data confirm this conception [29, 61] and give the
correlation length $T_g\simeq0.2 ~fm$.

The method of field correlators allows us to evaluate approximately the Wilson
loop using a cumulant expansion. MFC takes into account both perturbative and
nonperturbative interactions of gluons and therefore it is applicable for small and
large distances.

The Wilson integral is expressed via gauge-invariant
and Lorenz covariant correlators of strengths of gluonic fields. To make this expansion
one needs to use the non-Abelian Stokes theorem (25).

Let us introduce the generating functional for field correlators of gluonic fields
(see [65]) as follows
\begin{equation}
Z[J]=\langle \exp\left (ig\int dxJ_{\mu\nu}(x)F_{\mu\nu}(x,x_0)\right )\rangle,
\label{(138)}
\end{equation}
where $F_{\mu\nu}(x,x_0)=\Phi(x_0,x)F_{\mu\nu}(x)\Phi(x,x_0)$.
Using the Taylor expansion in $J_{\mu\nu}$ we arrive at
\[
Z[J]=1+\sum_{n=1}^{\infty}\frac {(ig)^n}{n!}\int dx_n\cdot\cdot\cdot
\int dx_n J_{\mu_1\nu_1}(x_1)\cdot \cdot \cdot J_{\mu_n\nu_n}(x_n)
\]
 \begin{equation}
\times\langle F_{\mu_1\nu_1}(x_1,x_0)\cdot \cdot \cdot F_{\mu_n\nu_n}(x_n,x_0)\rangle,
\label{(139)}
\end{equation}
so that the field correlators (averaged strength tensors over gluonic fields) are given by
\begin{equation}
\langle F_{\mu_1\nu_1}(x_1,x_0)\cdot \cdot \cdot F_{\mu_n\nu_n}(x_n,x_0)\rangle=
\frac{(ig)^{-n}\delta^n Z[J]}{\delta J_{\mu_1\nu_1}(x_1)\cdot \cdot \cdot
\delta J_{\mu_n\nu_n}(x_n)}\mid _{J=0}.
\label{(140)}
\end{equation}
It is convenient instead of expressions (140) to use irreducible correlators -
cumulants [64] in accordance with relationship
\begin{equation}
\langle \langle F_{\mu_1\nu_1}(x_1,x_0)\cdot \cdot \cdot F_{\mu_n\nu_n}(x_n,x_0)
\rangle \rangle =
\frac{(ig)^{-n}\delta^n \ln Z[J]}{\delta J_{\mu_1\nu_1}(x_1)\cdot \cdot \cdot
\delta J_{\mu_n\nu_n}(x_n)}\mid _{J=0}.
\label{141}
\end{equation}
From Eq. (141) we come to the expansion of the generating functional $Z[J]$ via
cumulants:
\[
Z[J]=\exp\biggl (1+\sum_{n=1}^{\infty}\frac {(ig)^n}{n!}\int dx_1\cdot\cdot\cdot
\int dx_n J_{\mu_1\nu_1}(x_1)\cdot \cdot \cdot J_{\mu_n\nu_n}(x_n)
\]
\begin{equation}
\times \langle \langle F_{\mu_1\nu_1}(x_1,x_0)\cdot \cdot \cdot
F_{\mu_n\nu_n}(x_n,x_0)\rangle \rangle \biggr ).
\label{142}
\end{equation}
Making variations of Eqs. (139) and (142) and setting $J_{\mu\nu}=0$ one arrives at
the first three relations [64] (we omit here argument $x_0$ implaying
$F_{\mu\nu}(x,x_0)=F_{\mu\nu}(x)$):
\[
\langle \langle F_{\mu\nu}(x)\rangle \rangle =\langle F_{\mu\nu}(x)\rangle,
\]
 \[
\langle \langle F_{\mu_1\nu_1}(x_1)F_{\mu_2\nu_2}(x_2)\rangle \rangle =
\langle F_{\mu_1\nu_1}(x_1)F_{\mu_2\nu_2}(x_2)\rangle -
\langle F_{\mu_1\nu_1}(x_1)\rangle \langle F_{\mu_2\nu_2}(x_2)\rangle,
\]
\[
\langle \langle F_{\mu_1\nu_1}(x_1)F_{\mu_2\nu_2}(x_2)F_{\mu_3\nu_3}(x_3)\rangle \rangle =
\langle F_{\mu_1\nu_1}(x_1)F_{\mu_2\nu_2}(x_2)F_{\mu_3\nu_3}(x_3)\rangle
\]
\[
-\langle F_{\mu_1\nu_1}(x_1)F_{\mu_2\nu_2}(x_2)\rangle \langle F_{\mu_3\nu_3}(x_3)\rangle
-\langle F_{\mu_1\nu_1}(x_1)\rangle \langle F_{\mu_2\nu_2}(x_2)F_{\mu_3\nu_3}(x_3)\rangle
\]
\[
-\langle F_{\mu_1\nu_1}(x_1)F_{\mu_3\nu_3}(x_3)\rangle\langle F_{\mu_2\nu_2}(x_2)\rangle
+\langle F_{\mu_1\nu_1}(x_1)\rangle \langle F_{\mu_2\nu_2}(x_2)\rangle
\langle F_{\mu_3\nu_3}(x_3)\rangle
\]
\begin{equation}
+\langle F_{\mu_1\nu_1}(x_1)\rangle\langle F_{\mu_3\nu_3}(x_3)\rangle
\langle F_{\mu_2\nu_2}(x_2)\rangle.
\label{143}
\end{equation}
It should be noted that there are two types of ordering in cumulant expansions [17], [64].
According to the definition of [17], there are no terms in Eq. (143)
like
\[
\langle F_{\mu_1\nu_1}(x_1)F_{\mu_3\nu_3}(x_3)\rangle\langle F_{\mu_2\nu_2}(x_2)\rangle,~
\langle F_{\mu_1\nu_1}(x_1)\rangle\langle F_{\mu_3\nu_3}(x_3)\rangle
\langle F_{\mu_2\nu_2}(x_2)\rangle,
\]
which violate path-ordering prescription.
For the Gaussian approximation when we retain only the bilocal cumulant both definitions coincide.
The average value $\langle F_{\mu\nu}(x)\rangle$ should vanish, because in the stochastic
vacuum there is no the definite direction. Otherwise the Lorentz
invariance will be broken by vacuum.
Applying the cumulant expansion (142) to the average Wilson integral (24) one arrives at
\[
\langle W(C)\rangle =\frac1{N_C} tr P \exp\biggl(\sum_{n=1}^{\infty}\frac {(ig)^n}{n!}
\int_{\Sigma}d\sigma_{\mu_1\nu_1}(x_1)\cdot\cdot\cdot
\int_{\Sigma} d\sigma_{\mu_n\nu_n}(x_n)
\]
\begin{equation}
\times \langle \langle F_{\mu_1\nu_1}(x_1,x_0)\cdot \cdot \cdot
F_{\mu_n\nu_n}(x_n,x_0)\rangle \rangle \biggr ),
\label{144}
\end{equation}
where $P$ an ordering operator on the coordinate $x$ of the matrix
$F_{\mu\nu}(x,x_0)$, and
we imply averaging on gluonic fields with the standard weight (see (5)).

 The expansion (144) makes sense when it converges. As we assume that Eq. (144) is
valid for any $g$ (even for large distances), cumulant expansion is nonperturbative.
Lattice data showed [29, 66] that the correlators decrease when the distance between two
points increases and the expansion (144) is justified.
It is known that in a Gaussian random process, the higher cumulants vanish, and only the
quadratic cumulant does not equal zero. With good accuracy the QCD vacuum can be considered
as a stochastic ensemble of gluonic fields [66]. This means that the main contribution to
the Wilson loop expansion (144) comes from the bilocal cumulant
\begin{equation}
\langle \langle F_{\mu\nu}(x,x_0)F_{\alpha\beta}(y,x_0)\rangle \rangle=
\langle \langle F_{\mu\nu}(x)\Phi(x,y)F_{\alpha\beta}(y)\Phi(y,x)\rangle \rangle.
\label{145}
\end{equation}
 Higher cumulants in Eq. (144) give small corrections with
$\sqrt {\langle (F_{\mu\nu}^a)^2 \rangle}T_g^2\ll 1$ [16, 17], and can be neglected
(Gaussian approximation). The Gaussian dominance was supported by lattice calculations
and the hadron phenomenology. Thus we come to the Gaussian stochastic model of the QCD vacuum
[29, 66]. The fundamental input in such a model is the Gaussian correlator. But there is a
problem; what kind of field configurations (classical or quantum) play the dominant role to
form the definite behavior of field correlators.

However, there will be the dependence of Wilson loop
integral (144) on the point $x_0$ and the shape of the surface $\Sigma$. We recall
that the point $x_0$ and contour $C$ belong to the surface $\Sigma$. The total expression
(144) with higher cumulants does not depend on the point $x_0$ and contour $C$.
To reduce the dependence of a cumulant on the $\Sigma$, one should choose the surface
with the minimal area [17]. Then the area law of the Wilson integral with large loops
(the size $R\simeq 1 ~fm$) takes place [17]. To get rid on the dependence of the Wilson
loop on $x_0$ one should have the condition [17]
\[
\mid x^{(i)}-x^{(j)}\mid\ll\mid x^{(i)}-x_0\mid,~~~ \mid x^{(j)}-x_0\mid.
\]
Thus, at small correlation length $T_g\ll\mid x^{(i)}-x_0\mid$ one can neglect the
dependence on $x_0$.
In the Gaussian approximation the Wilson integral (144) becomes
\[
\langle W(C)\rangle =\frac1{N_C} trP\exp\biggl(-\frac{g^2}{2}
\int_{\Sigma_{min}}d\sigma_{\mu\nu}(x)\int_{\Sigma_{min}} d\sigma_{\alpha\beta}(y)
\]
\begin{equation}
\times \langle \langle F_{\mu\nu}(x,x_0)F_{\alpha\beta}(y,x_0)\rangle \rangle \biggr ),
\label{146}
\end{equation}
where we imply that $x=x(\xi)$, $y=y(\xi)$.
 Thus, for a size of the Wilson loop $R\gg T_g$ the bilocal
 cumulant (145) entering Eq. (146) does not depend on $x_0$. As a result it can be expressed
via two scalar functions $D_1$ and $D_2$ [16,17]:
\[
\frac{g^2}{2}\langle \langle F_{\mu\nu}(x)\Phi(x,y)F_{\alpha\beta}(y)\Phi(y,x)\rangle \rangle
=I_{N_C}\biggl\{(\delta_{\mu\alpha}\delta_{\nu\beta}-\delta_{\mu\beta}\delta_{\nu\alpha})D\left(
x-y\right)
\]
\[
+\frac12\biggl[\frac{\partial}{\partial x_\mu}\biggl( (x-y)_\alpha\delta_{\nu\beta}-
(x-y)_\beta \delta_{\nu\alpha}\biggr) \]
\begin{equation}
+\frac{\partial}{\partial x_\nu}\biggl( (x-y)_\beta
\delta_{\mu\alpha}
-(x-y)_\alpha \delta_{\mu\beta}\biggr)\biggr]D_1\left(x-y\right)\biggr \},
\label{147}
\end{equation}
where $I_{N_C}$ is a unit matrix $N_C\times N_C$, and functions $D$ and $D_1$ are invariants
of the renormalization group. As the cumulant (147) is proportional to the unit matrix $I_{N_C}$,
the ordering operator $P$ can be omitted in Eq. (146).
Substituting Eq. (147) into Eq. (146) one arrives at Eg. (29) with the string tension [17]:
\begin{equation}
\sigma=\frac12\int d^2xD(x).
\label{148}
\end{equation}
So, $\sigma$ is an integral characteristic of the Gaussian correlator.
Only function $D$ enters Eq. (148) defining the string tension but both functions $D$ and
$D_1$ give the contribution
to the perimeter term $\gamma L(C)$ in Eq. (31) [17]. Higher cumulants add small corrections
to the string tension (148). The functions $D, D_1$ in the nonperturbative regime were
derived in lattice experiments [29,66] and are given by
\begin{equation}
D(x)=D(0)\exp \biggl(-\frac{\mid x\mid}{T_g}\biggr), ~~
D_1(x)=D_1(0)\exp \biggl(-\frac{\mid x\mid}{T_g}\biggr),
\label{149}
\end{equation}
where $D(0)=0.073 ~GeV^4$, $D_1(0)=(1/3)D(0)$, $T_g=0.2\div 0.3 ~fm$. Values $D(0), D_1(0)$ can
be connected with the gluonic condensate [4]
\begin{equation}
\frac{\alpha_s}{\pi}\langle \left(F^a_{\mu\nu}(0)\right)^2\rangle\simeq 0.012 ~GeV^4,
\label{150}
\end{equation}
so that $D(0)\simeq 2\alpha_s\langle \left(F^a_{\mu\nu}(0)\right)^2\rangle$. Thus,
MFC is a development of the QCD sum rule method.
In the perturbative regime at small distances, the contribution from gluonic exchange
(giving the Coulomb potential) to the $D(x), D_1(x)$ is of $O(g^2)$ [17]
\begin{equation}
D(x)=0,~~~ D_1(x)=\frac{16\alpha_s}{3\pi x^4}.
\label{151}
\end{equation}
The colour Coulomb interaction is dominant below $0.3 ~fm$. This region is important for
bottomonium ($\bar {b}b$ state) and charmonium ($\bar {c}c$ state) where the sizes of ground
states are around $0.2 ~fm$ and $0.4 ~fm$, respectively. For $\bar {c}c$ and $\bar {s}s$
systems both perturbative and nonperturbative interactions should be taken
into account, and only two $D$, $D_1$ functions can be used. It should be noticed that
nonperturbative sector influences the perturbative part and vice versa. So, two functions,
$D$, $D_1$ can be represented as [17,66]
\begin{equation}
D(x)=D^{Pert}(x)+D^{NP}(x),~~~ D_1(x)=D_1^{Pert}(x)+D_1^{NP}(x).
\label{152}
\end{equation}
At $x^2\rightarrow 0$ the perturbative part of the function $D$ has the form $D^{Pert}(x)
\sim x^{-\alpha}$, $\alpha > 0$. The nonperturbative part is normalized by the gluon
condensate
\[D^{NP}(0)+D_1^{NP}(0)=(1/24N_C)\langle \left(F^a_{\mu\nu}(0)\right)^2
\rangle=(\pi^2/6N_C) (0.14\pm0.02) ~GeV^4.
\]
The value of the gluon condensate was taken from the lattice data.

The strong nonperturbative chromoelectromagnetic fields of the QCD vacuum cause
nonperturbative shift of the energy density of the vacuum due to the scale anomaly [4]:
\begin{equation}
\epsilon=\frac{\beta(\alpha_s)}{16\alpha_s}\langle(F_{\mu\nu}(0)^a)^2\rangle.
\label{153}
\end{equation}
We know that the Gell-Mann - Low function, $\beta(\alpha_s)<0$ due to asymptotic freedom at
small $\alpha_s$. So,
the nonperturbative shift of the vacuum energy density is energetically favorable.

Because only function $D(x)$ contributes to the string tension (148), it is responsible
for a condensation of monopoles in accordance with the 't Hooft and Mandelstam scenario of
confinement. The lattice calculations gave the width of the string $l=0.4\div0.5 ~fm
$, so that the correlation length $T_g\simeq (1/2)l$. Thus, the radius of the string is
close to $T_g$. Therefore another interpretation of $T_g$ is the thickness of the
confining string. In the case $T_g\rightarrow 0$ we have pure string limit.
The reader can find more about MFC in [17,66].

\section{Conclusion}

The QCD string theory allows us to obtain the effective Lagrangians for mesons
and baryons and estimate the mean squared radii and electromagnetic
polarizabilities of hadrons. These quantities were derived as
functions of the string tension which is a fundamental variable in this
approach. It is not difficult to calculate the electromagnetic polarizabilities of
excited states of hadrons using this approach. For that we should take the
quantum numbers $n_r=1$, $l=0$ ~$(n=2)$ and evaluate the mean relative
coordinate in accordance with Eq. (113). Then Eqs. (118), (130) give the required
polarizabilities. To have more precise values of the hadron electromagnetic
characteristics one needs to take into account spin corrections. Especially
it is important for light pseudoscalar mesons ($\pi$, $K$ mesons). In principle
it is possible to receive the spin-orbit and spin-spin interactions using the
general expressions (23), (83). The model of baryons as a quark-diquark system
is very similar to the approach for mesons. The mean size and electromagnetic
polarizabilities of a proton and neutron are in reasonable agreement with the
experimental data. The more favorable combination for a diquark is $(uu)$ for
a proton and $(dd)$ for a neutron. In this case theoretical values for electromagnetic
polarizabilities are close to experimental data.

It should be noted that the quark-diquark ansatz used here generates an electric dipole
moment, and a result, there is a contribution to the energy linear in the electric field
which is like the linear Stark effect. This requires the deeper understanding the properties
of a nucleon under the symmetry transformations (for example parity).

Spin interactions of quarks treated as a perturbation were neglected here but we took them
into account by using the paramagnetic polarizability of a nucleon due to the
$\Delta$ contribution. As an approximation it is justified because in Isgur-Karl model
of baryons [58] spin-orbit splitting are much smaller than expected from one-gluon-exchange
matrix elements (spin forces were also discussed in [59]). It was also shown [60,61]
that the contributions from the Coulomb and spin-spin interactions cancel each other.
Nevertheless all parameters should be taken from the approach and then compared with
empirical data. Therefore one needs to calculate the paramagnetic polarizability of a nucleon
in our approach.

Implying small spin-orbit forces in baryons we come to $K^{*}$ and quark-diquark
system correspondence (see [36]). The present approach can be applied for
studying any baryon (see also [60]).

The Nambu-Jona-Lasinio (NJL) model [13,14] having a basis in the framework
of QCD [12] describes the chiral symmetry breaking but not the confinement
of quarks [17,15]. Besides this model has free parameters and the calculated
polarizabilities of mesons [62] are parameter dependent.

The instanton vacuum theory developed in [11] does not give the
confinement of quarks phenomenon [17]. This theory is like the NJL model [15]
and takes into account only the chiral symmetry breaking. Therefore the
calculation of the meson electromagnetic polarizabilities on the basis of the
IVT gave the similar result [63] as in the NJL model.

The electromagnetic polarizabilities of nucleons found are close to the values
calculated in the framework of the dispersion sum rule [71] and the chiral
perturbation theory (CHPT) [72].

To improve the accuracy of calculations of electromagnetic polarizabilities of
nucleons one should take into account pion degrees of freedom because the pion
cloud contributes substantially to electromagnetic properties (see Appendix).

All this shows that the theoretical evaluation of the charge radii and the
magnetic polarizabilities of hadrons is possible on the basis of
a good description the chiral symmetry breaking and confinement of quarks in
the framework of the QCD string theory but with some approximations and
model assumptions. Naturally this theory was derived using the
non-perturbative QCD, i.e. first principals of QCD.

\section{Appendix}
In this Appendix we define nucleon electromagnetic polarizabilities in different approaches.
The static electric $\alpha$ and magnetic $\beta$ polarizabilities of a hadron characterize
its internal structure. These quantities are low-energy characteristics of a hadron and therefore
they are defined in the non-perturbative regime of QCD. In external electromagnetic fields
the hadron is deformed and induces dipole electric and magnetic moments
\begin{equation}
{\bf D} = \alpha {\bf E},~~~{\bf M} = \beta {\bf H}.
\label{154}
\end{equation}
The acquired dipole moments (154) of a hadron give the contribution to the potential
energy as follows
\begin{equation}
U(\alpha ,\beta )=-\frac 12\alpha E^2-\frac 12\beta H^2.
\label{(155)}
\end{equation}
We use Gaussian units here.
The cross section of scattering of a photon on hadrons (Compton scattering) depends on the
electromagnetic polarizabilities due to Eq. (155). Thus, low-energy Compton scattering
can give important information about the internal structure of hadrons.
Below we discuss the nucleon electromagnetic polarizabilities. The Compton amplitude
for scattering a photon on a nucleon at low energies in the nucleon rest frame (the laboratory
system) is given by [22]
\begin{equation}
f(\gamma N \rightarrow \gamma N)=f_B+\omega \omega'\bar{\alpha} ({\bf \epsilon}
{\bf \epsilon}')+ \bar{\beta} ({\bf k}\times {\bf \epsilon})
({\bf k}'\times {\bf \epsilon}')+O(\omega^3),
\label{(156)}
\end{equation}
where $({\bf k},i\omega)$, ${\bf \epsilon}$ and $({\bf k}',i\omega')$,
${\bf \epsilon}'$ are four momenta and polarization vectors of the incoming
and outgoing photon, respectively. The electromagnetic polarizabilities $\bar{\alpha}$,
$\bar{\beta}$ occurring Eq. (156) are so-called Compton polarizabilities. Below we will
discuss the interrelation between static and Compton polarizabilities.
The $f_B$ is the Born amplitude which is due to
the nucleon electric charge $eZ$ and anomalous magnetic moment $\kappa$ of a nucleon.
The Born amplitude $f_B$ includes the Thomson amplitude $f_T$ which is energy-independent
and satisfies the well-known low-energy theorem:
\begin{equation}
f_T=-\frac{e^2Z^2}{4\pi M_N}({\bf \epsilon}{\bf \epsilon}'),
\label{(157)}
\end{equation}
where $M_N$ is the nucleon mass. The Thomson scattering amplitude (157) corresponds
to a pointlike particle and the total amplitude (156) includes structure parameters -
electromagnetic polarizabilities $\bar{\alpha}$, $\bar{\beta}$.

In the nonrelativistic approximation the generalized electric and magnetic polarizabilities
which are extracted from measurements of the Compton scattering cross sections are given by [22]
\[
\bar{\alpha }=\alpha +\Delta \alpha,
\]
\begin{equation}
\alpha =2\sum_{n\ne N} \frac{\mid \langle n\mid D_z\mid
N\rangle \mid ^2}{{\cal E}_n-{\cal E}_0}.
\label{(158)}
\end{equation}
 \[
\bar{\beta }=\beta ^{para}+\beta ^{dia},
\]
\begin{equation}
\beta ^{para}=2\sum_{n\ne N} \frac{\mid \langle n\mid M_z\mid
N\rangle \mid ^2}{{\cal E}_n-{\cal E}_0}.
\label{(159)}
\end{equation}
where $\mid n\rangle $ is the excited state of
a hadron, $D_z$ and $M_z$ are the third projections of the electric and magnetic
dipole operators, respectively. The  excited state $\mid n\rangle$ includes the meson-baryon
intermediate states. In nonrelativistic calculations the $\Delta \alpha $ is due to form factors
and depends on the electric charge radius.
Eqs. (158), (159) are similar to the usual quantum-mechanical formulas.
Eq. (158) was also confirmed by the quasiclassical calculations of pion polarizability in
the framework of the instanton vacuum theory [63]. In the completely relativistic approach the
physical meaning of $\bar{\alpha}$ is more complicated [22].
In [22] the correction $\Delta \alpha $ is given by
\begin{equation}
\Delta \alpha =\frac{e^2r_E^2}{3M_N}+\frac{e^2\left( \kappa^2 +1\right) }{4M_N^3},
\label{(160)}
\end{equation}
where $r_E$ is the electric radius and the magnetic moment of the hadron is given by
$\mu =(e\kappa )/(2M_N)$.

For the proton and neutron, using the experimental values
of electric radii and magnetic moments, the term (160) becomes [22]
\begin{equation}
\Delta \alpha_p=(4.5\pm 0.1)\times 10^{-4}~ fm^3,~~~
\Delta \alpha_n=0.62\times 10^{-4}~ fm^3
 \label{(161)}
\end{equation}
The strong magnetic $N\Delta$ transition gives rise to a positive part of the
magnetic polarizabilities. Therefore the main contribution to the paramagnetic polarizability
of a nucleon is due to $\Delta (1232)$ excitation [57] and is given by
\begin{equation}
\beta _\Delta
^{para}=\left( 13\pm 3\right) \times 10^{-4}~ fm^3.
\label{(162)}
\end{equation}
The intermediate state $\Delta$ is an isovector excitation and contributes
to the proton and neutron so that $\beta _\Delta ^{p}=\beta _\Delta ^{n}$.
The interesting feature of the experimental data is that for the proton and the neutron
the approximate equality $\bar {\alpha}_p+\bar {\beta}_p \simeq \bar {\alpha}_n+
\bar {\beta}_n$, and inequality $\bar {\alpha}_p, \bar {\alpha}_n\gg
\bar {\beta}_p, \bar {\beta}_n$ are valid. The last inequality means that both
a proton and a neutron behave like electric dipoles. In addition, the large positive
paramagnetic polarizability of nucleons should be canceled by the diamagnetic contribution.
The sum of electromagnetic polarizabilities can be connected with the total photoabsorption cross
section [67,68,22]. Indeed, from Eq. (156) the spin-independent amplitude
of the Compton scattering at low energies and zero angle and ${\bf \epsilon}={\bf \epsilon}'$
is given by
\begin{equation}
f_{\gamma \rightarrow \gamma'}(\omega,0)= -\frac{e^2Z^2}{4\pi M_N}+ (\bar{\alpha} +
\bar{\beta}) \omega ^2 +O(\omega^3).
\label{(163)}
\end{equation}
Using the optical theorem [22]
\begin{equation}
Im f_{\gamma \rightarrow \gamma'}(\omega,0)= \frac{\omega}{4\pi}\sigma_{tot}(\omega),
\label{(164)}
\end{equation}
and once-subtracted forward dispersion relation for amplitude (163), one finds
the relation [67,68]
\begin{equation}
(\bar{\alpha} + \bar{\beta})= \frac{1}{2\pi ^2}\int_{\omega_{thr}}^{\infty}
\frac{d\omega}{\omega^2}\sigma_{tot}(\omega),
\label{(165)}
\end{equation}
where $\omega=(s-M_N^2)/(2M_N)$ is the photon energy in the laboratory system and $\omega_{thr}=
m_\pi(1+m_\pi/(2M_N))$ is the pion production threshold ($m_{\pi}=139.57 ~MeV$ is the
pion mass). Eq. (165) occurs also for spin-0 particles. The strict equality (165) allows
us to extract the sum of electromagnetic polarizabilities from the experimentally measured
total photoabsorption cross section. The result is [69,70]
\[
(\bar{\alpha}_p + \bar{\beta}_p)=\left(14.3\pm 0.5\right) \times 10^{-4}~ fm^3,
\]
\begin{equation}
(\bar{\alpha}_n + \bar{\beta}_n)=\left(15.8\pm 0.5\right) \times 10^{-4}~ fm^3.
\label{(166)}
\end{equation}
The most difficult task is to find electromagnetic polarizabilities separately (see [22]).

It follows from Eq. (165) that at energies $\omega\simeq m_{\pi}$ the sum of polarizabilities
behave as $O(m_{\pi}^{-1})$ because the total photoabsorbtion
cross section $\sigma_{tot}(\omega)$ is finite at $\omega\simeq m_{\pi}\rightarrow 0$ [71].
In the chiral limit, $m_{\pi}\rightarrow 0$, both polarizabilities $\bar{\alpha}$, $\bar{\beta}$
diverge as $O(m_{\pi}^{-1})$. This was confirmed in the framework of the chiral perturbation
theory (CHPT) [72]. The same behavior of pion polarizabilities was found on the basis of
quasiclassical calculations [63] in the framework of the instanton vacuum theory.
Thus, the leading terms of electromagnetic polarizabilities are determined by CSB.

The contribution of low-lying intermediate states, where soft pions play
an important role, into the integral of Eq. (165) was estimated in [71]:
\[
\bar{\alpha}= \frac{5}{12m_\pi}\left(\frac{eg_{\pi N}\sqrt 2}{8\pi M_N}\right)^2\simeq
 13.6\times 10^{-4}~ fm^3,
\]
\begin{equation}
\bar{\beta} =\frac{1}{24m_\pi}\left(\frac{eg_{\pi N}\sqrt 2}{8\pi M_N}\right)^2\simeq
 1.4\times 10^{-4}~ fm^3,
\label{(167)}
\end{equation}
where the strong pion-nucleon coupling constant $g_{\pi N}=13.4$ is connected with the
charged pion decay constant $f_{\pi}=93 ~MeV$ by the Goldberger-Treiman relation $g_{\pi N}/
M_N=g_A/f_\pi$, where $g_A$ is the axial-vector coupling constant. The values (167) found
for the leading contribution in the chiral limit, and
therefore $\bar{\alpha}$, $\bar{\beta}$ are the same for the proton and neutron. The next
corrections $O(\omega/M_N)$ to polarizabilities, due to the nucleon recoil, are different
for a proton and neutron [71]. The result (167) is in agreement with CHPT calculations [72].

The nucleon polarizabilities were also evaluated on the basis of nucleon soliton
models in [73-79] .

\end{document}